\documentclass[a4paper,aps,onecolumn,nofootinbib,10pt]{revtex4}

\usepackage{float}
\usepackage{graphicx}
\usepackage{caption}  
\usepackage{slashed}
\usepackage[english,]{babel}
\usepackage{braket}
\usepackage{enumitem}
\usepackage{makeidx,xspace}
\usepackage{epsfig,amsfonts,amssymb,amsmath}
\usepackage{cancel}
\usepackage{mathrsfs}
\usepackage{extsizes}
\usepackage[a4paper,top=3cm,bottom=3cm,left=2cm,right=2cm]{geometry}
\usepackage[colorlinks,hyperindex]{hyperref}
\hypersetup{colorlinks=true,breaklinks=true,urlcolor=blue,linkcolor=red}
\makeindex

\newcommand{\be}{\begin{equation}}\newcommand{\ee}{\end{equation}}
\newcommand{\bea}{\begin{eqnarray}}\newcommand{\eea}{\end{eqnarray}}
\newcommand{\brr}{\begin{array}}\newcommand{\err}{\end{array}}
\newcommand{\bit}{\begin{itemize}}\newcommand{\eit}{\end{itemize}}
\newcommand{\ben}{\begin{enumerate}}\newcommand{\een}{\end{enumerate}}

\def\1{{_{1}}}\def\2{{_{2}}}

\begin{document}
\title{A covariant approach to the Dirac field in LRS space-times: the case of coplanar frames}
\author{Stefano Vignolo,$^{c}$\! $^{G}$\footnote{stefano.vignolo@unige.it}Giuseppe De Maria,$^{c}$\! $^{\hbar}$\footnote{giuseppe.demaria@edu.unige.it}Sante Carloni,$^{c}$\! $^{\hbar}$\! $^{G}$\footnote{sante.carloni@unige.it}Luca Fabbri,$^{c}$\! $^{\hbar}$\! $^{G}$\footnote{luca.fabbri@unige.it}}
\affiliation{$^{c}$DIME, Universit\`{a} di Genova, Via all'Opera Pia 15, 16145 Genova, ITALY\\
$^{\hbar}$INFN, Sezione di Genova, Via Dodecaneso 33, 16146 Genova, ITALY\\
$^{G}$GNFM, Istituto Nazionale di Alta Matematica, P.le Aldo Moro 5, 00185 Roma, ITALY\\}
\date{\today}
\begin{abstract}
We employ the polar decomposition of the Dirac field to describe it as an effective spinorial fluid. We then construct a 
$(1+1+2)$ covariant formalism for the Dirac field that avoids the introduction of tetrad fields and Clifford matrices. Within this framework, we analyze the conditions under which a self-gravitating Dirac field can be consistently embedded in Locally Rotationally Symmetric (LRS) space-times of types I, II, and III. In accordance with the LRS symmetry requirements, we extend a previous work by assuming that the velocity and spin vector fields of the Dirac field lie in the planes defined pointwise by the generators of the time-like and space-like congruences, which underlie the 
$(1+1+2)$ decomposition. We present some analytical and numerical solutions to illustrate the applicability of the proposed framework.
\end{abstract}
\maketitle
\section{Introduction}\label{Section1}
The covariant approach to general relativity, originally developed by Ehlers \cite{Ehlers:1961xww} and later systematized by Ellis and collaborators \cite{Ellis1971,cargese,Ellisperf,Ellis1971bis}, provides an extraordinarily powerful geometric tool for describing relativistic space-times. By formulating the dynamics in terms of covariantly defined quantities relative to a given time-like congruence, the formalism avoids reliance on specific coordinate systems and yields a description directly tied to physically measurable variables. This makes it particularly suitable for applications in cosmology and astrophysics, where the choice of observers plays a central role.

From its first formulation to the present day, the covariant formalism has been widely applied to a broad range of problems in gravitational physics, including anisotropic and inhomogeneous cosmologies, cosmological perturbation theory, relativistic magnetohydrodynamics, black holes, relativistic stars and gravitational collapse \cite{eb,ehb,ebh,bed,1995ApJ...443....1S,Maartens1998,Clarkson,Tsagas,Clarkson2,Ellis2011,Umeh,Umeh2,Carloni:2017rpu,Carloni:2017bck,Naidu:2021nwh,Naidu:2022igk,Luz:2024yjm,Luz:2024lgi,Luz:2024xnd}. 

The covariant formalism relies primarily on the ability to describe matter fields as effective fluids. It therefore applies naturally to standard relativistic fluid continua and has also been successfully extended to other types of matter fields, such as scalar fields \cite{Carloni:2006gy,Carloni:2019cyo}. In covariant approaches to relativistic theories, the key point is to express the matter energy-momentum tensor in the same form as that of a fluid. This aspect, however, represents one of the main challenges in developing covariant approaches to the Dirac field. Indeed, in the standard treatment of spinor fields, the presence of Clifford matrices, the nontrivial spinorial derivative, and the specific choice of a tetrad field make the covariant approach difficult to apply.

One possible way to overcome this difficulty is provided by the so-called polar formalism for the Dirac field \cite{jl1,jl2,tr1,tr2,t2,Fabbri:2023onb,Fabbri:2023dgv}. In the polar form, the spinor field is described in terms of a set of real variables, typically including a scalar density (the module), a pseudo-scalar (the chiral angle), a time-like vector field (the velocity), and a space-like vector field (the spin), together with a phase parameter. This decomposition allows one to rewrite the Dirac equation in a manifestly covariant form involving only real tensorial quantities, thereby providing a clearer geometrical and physical interpretation of the spinor degrees of freedom.

Among its several advantages, the polar formalism allows for a genuinely hydrodynamic formulation of the Dirac field, providing a representation of its energy-momentum tensor in terms of that of an effective fluid \cite{Fabbri:2025ffi}. This enables a direct application of the covariant approach, avoiding any use of Clifford matrices and the need to choose specific tetrads for the soldering.

A first step in this direction was taken in \cite{VDFC}, where a covariant formulation of a self-gravitating Dirac field in locally rotationally symmetric (LRS) space-times was developed without resorting to the tetrad formalism. In particular, it was shown that the velocity and spin (pseudo) vector fields naturally define the time-like and space-like congruences that give rise to the $(1+1+2)$ covariant splitting and its associated geometrical framework.

Although natural at first glance, the identification of the two geometric congruences with the integral curves of the velocity and spin fields is somewhat restrictive and may be responsible for some of the obstructions encountered in our previous analysis. For instance, under the assumptions adopted in \cite{VDFC}, it was shown that no LRS-III solutions exist.

In this work, we extend the geometric construction proposed in \cite{VDFC} by investigating the compatibility of the Dirac field with LRS geometries of types I, II, and III in the most general setting, with the aim of identifying a broader class of solutions than those found in \cite{VDFC}. The new framework is obtained by dispensing with the identification of velocity and spin as generators of the two geometric congruences and instead requiring that they be coplanar with the unit tangent vectors to the two congruences. This condition turns out to be the most general choice compatible with the requirements of LRS geometry.

Within the resulting geometrical setting, after expressing both the energy-momentum tensor and the Dirac equation in polar form, we carry out their $(1+1+2)$ decomposition and derive the covariant equations for a Dirac field with backreaction in LRS space-times of types I, II, and III. As expected, these equations are generally strongly coupled and difficult to solve. We obtain analytical solutions in the simpler case of a homogeneous and isotropic space-time. In more complex scenarios, we investigate the physical properties of the solutions through numerical analysis.

This paper is organized as follows. For the reader's convenience, Section \ref{Section2} summarises the main features of the $(1+1+2)$ splitting in signature $(+---)$, commonly adopted in the treatment of spinor fields. Section \ref{Section3} reviews the basic aspects of the polar formulation of the Dirac field. Sections \ref{Section4} and \ref{CsApp} implement the $(1+1+2)$ covariant decomposition of the polar formalism extending that proposed in \cite{VDFC}. Section \ref{Section2bis} presents the main aspects of LRS geometries, again in signature $(+---)$. Section \ref{Section5} combines the polar formalism with the covariant approach to develop a more general covariant formulation of a self-gravitating Dirac field in LRS space-times of types I, II, and III. Section \ref{sezione_8} presents some exact and numerical solutions. Section \ref{Conclusion} is devoted to conclusions. 

Throughout the paper, we adopt natural units ($c=8\pi G=\hbar=1$) and metric signature $(+---)$. We write Einstein's equations (with cosmological constant $\Lambda=0$) as
$$
G_{ab}=T_{ab}
$$ 
where $G_{ab}$ and $T_{ab}$ are the Einstein and the energy--momentum tensors respectively. We express the Riemann tensor as
$$
R^{a}{}_{bcd} := \partial_{c}\Gamma_{db}{}^{a} - \partial_{d}\Gamma_{cb}{}^{a} + \Gamma_{cp}{}^{a}\Gamma_{db}{}^{p} - \Gamma_{dp}{}^{a}\Gamma_{cb}{}^{p}
$$
where $\Gamma_{ab}{}^{c}\partial_c:=\nabla_{\partial_a}\partial_b$, $\nabla$ denoting the (Levi--Civita) covariant derivative. We define the Ricci tensor as $R_{ab}:=R^{c}{}_{acb}$. For the symmetrization and antisymmetrization of expressions with two indexes we use the convention $W_{(ab)}:=\frac{1}{2}\left(W_{ab}+W_{ba}\right)$ and $W_{[ab]}:=\frac{1}{2}\left(W_{ab}-W_{ba}\right)$. Moreover, we use the following convention for the Levi-Civita tensor
$$
\varepsilon_{0123}=\sqrt{-g}, \qquad \varepsilon^{0123}=-\tfrac{1}{\sqrt{-g}}
$$
where $ g$ is the determinant of the metric tensor $ g_{ab} $.

\section{The (1+1+2) covariant approach in signature (+,--,--,--)}\label{Section2}
The $(1+1+2)$ covariant approach is based on the simultaneous assignment of two mutually orthogonal congruences, one time-like and the other space-like. Denoting respectively by $v^i$ and $e^i$ the unit vector fields tangent to the given congruences, we have the relations
\begin{equation}
v_iv^i = 1, \qquad e_ie^i = -1 \qquad {\rm and} \qquad e^iv_i = 0.
\end{equation}
At each point of space-time, the corresponding tangent space is decomposed into the direct sum of a subspace generated by $v^{i}$ and a three-dimensional subspace orthogonal to $v^i$. The latter is, in turn, decomposed into two orthogonal parts: a one-dimensional subspace generated by $e^i$ and a two-dimensional subspace orthogonal to both $v^i$ and $e^i$. Within this general framework, the metric tensor $g_{ij}$ takes the form
\begin{equation}\label{metric_112}
g_{ij} = v_i v_j + h_{ij} \qquad {\rm with} \qquad h_{ij}:= - e_i e_j + N_{ij},
\end{equation}
where $N_{ij}$ represents the induced metric on the two-dimensional subspaces orthogonal to $v^i$ and $e^i$. Instead, $h_{ij}$ is the induced metric on the three-dimensional subspaces orthogonal to $v^i$. By construction, the tensors $h_{ij}$ and $N_{ij}$ satisfy the conditions
\begin{subequations}\label{condizioni_h_N}
\begin{align}
&h_{ij}v^j = 0, \qquad h_{ij}h^j{}_h = h_{ih}, \qquad h^i{}_i = 3,\label{condizioni_h}\\
&N_{ij}v^j = 0, \qquad N_{ij}e^j = 0, \qquad N_{ij}N^j{}_h = N_{ih}, \qquad N^i{}_i = 2\label{condizioni_N}.
\end{align}
\end{subequations}
Any spatial vector $V^i$ ($V^i v_i = 0$) can be expressed as
\begin{equation}\label{vector_decomposition}
V^i = -V\/e^i + \mathcal{V}^i, \qquad {\rm with} \qquad
V := V^i\/e_i \qquad {\rm and} \qquad \mathcal{V}^i := N^i{}_j V^j.
\end{equation}
Similarly, any projected symmetric trace-free (PSTF) tensor $W^{ab}=W^{\langle ab \rangle}$ of ${\rm rank}=2$, where
\be
W^{\langle ab \rangle}:=\left[h^{(a}\hspace{0.1mm}_{c}h^{b)}\hspace{0.1mm}_{d}-\frac{1}{3}h^{ab}h_{cd} \right]W^{cd}, 
\ee
can be written in irreducible parts as 
\begin{equation}\label{tensor_decomposition}
W^{ab} = W \left( e^a\/e^b + \frac{1}{2}N^{ab} \right)
 - 2 \mathcal{W}^{(a}\/e^{b)} + \mathcal{W}^{ab},
\end{equation}
where
\begin{align}
W &= W^{ab} e_a e_b = W^{ab} N_{ab }, \\
\mathcal{W}^a &= N^a{}_b e_c W^{bc}, \\
\mathcal{W}^{ab} &= \left(N^{(a}{}_c\/N^{b)}{}_d - \frac{1}{2} N^{ab}\/N_{cd}\right)W^{cd}.
\end{align}
Denoting by $\varepsilon_{abcd}$ the Levi-Civita tensor, we introduce the alternating tensor
\begin{equation}\label{alternating_tensor}
\varepsilon_{ab} := \varepsilon_{jab}e^j := \varepsilon_{ijab} v^i e^j,
\end{equation}
which satisfies the relation
\begin{equation}
\varepsilon_{ab} \varepsilon^{cd} = N_a{}^c N_b{}^d - N_a{}^d N_b{}^c .
\end{equation}
For every tensor field $A^{a}\hspace{0.1mm}_{...b}$, a covariant {\it time} derivative 
\bea\label{time_cov_der}
\dot{A}^{a}\hspace{0.1mm}_{...b}:= v^{c}\nabla_{c}A^{a}\hspace{0.1mm}_{...b} ,
\eea 
as well as a fully orthogonally projected (compared to $v^i$) covariant derivative 
\bea\label{space_cov_der} 
D_{c}A^{a}\hspace{0.1mm}_{...b}:= h^{d}\hspace{0.1mm}_{c}h^{a}\hspace{0.1mm}_{f}...h^{g}\hspace{0.1mm}_{b}\nabla_{d}A^{f}\hspace{0.1mm}_{...g} 
\eea
are defined. On the basis of the spatial covariant derivative \eqref{space_cov_der}, two further covariant derivatives
\begin{align}
\hat{A}^a{}_{\dots b} &:= e^c D_c A^a{}_{\dots b}, \label{hat_derivative}\\
\delta_c A^a{}_{\dots b} &:= N_c{}^d N^a{}_e \dots N^f{}_b D_d A^e{}_{\dots f} \label{delta_derivative}
\end{align}
can be introduced.
In view of this, the covariant derivatives of the unit vector fields $v^i$ and $e^i$ can be expressed as \cite{VDFC}
\begin{subequations}\label{covariant_derivatives_un}
\begin{align}
\nabla_a v_b &= -v_a(A e_b - \mathcal{A}_b) + e_a e_b \!\left(\Sigma - \tfrac{1}{3}\Theta\right)
- e_a(\Sigma_b - \varepsilon_{bc}\Omega^c)
- e_b(\Sigma_a + \varepsilon_{ac}\Omega^c) \nonumber\\
&\quad + N_{ab}\!\left(\tfrac{1}{3}\Theta + \tfrac{1}{2}\Sigma\right)
+ \Omega \varepsilon_{ab} + \Sigma_{ab}, \label{cdu_final}\\[2mm]
\nabla_a e_b &= -A v_a v_b + v_a \alpha_b + \left(\Sigma - \tfrac{1}{3}\Theta\right) e_a v_b 
- (\Sigma_a + \varepsilon_{ac}\Omega^c)v_b - e_a a_b \nonumber\\
&\quad + \tfrac{1}{2}\phi N_{ab} + \xi \varepsilon_{ab} + \zeta_{ab}, \label{cdn_final}
\end{align}
\end{subequations}
where the following quantities are involved:
\begin{itemize}
\item scalar components
\begin{eqnarray}\label{kinematical_quantities scalars}
\nonumber&A = e^{a}v^{i}\nabla_{i}v_{a}, \qquad
\Sigma=\left[\frac{1}{2}(D_a v_b+D_b v_a)-\frac{1}{3}D_k\/v^k h_{ab}\right]e^a\/e^b, \qquad
\Omega = \frac{1}{2}\varepsilon^{ba}\nabla_{b}v_{a}, \\[2mm]
&\phi=N^{ab}\nabla_{a}e_{b}, \qquad
\Theta=\nabla_a\/v^a, \qquad
\xi=\frac{1}{2}\varepsilon^{ba}\nabla_{b}e_{a}; 
\end{eqnarray}
\item vector components
\begin{eqnarray}\label{kinematical_quantities vectors}
\nonumber &\mathcal{A}^{b}=N^{ab}v^{i}\nabla_{i}v_{a}, \qquad
\alpha^{b}=N^{bk}v^{i}\nabla_{i}e_{k}, \qquad
a^{b}=N^{bk}e^{i}\nabla_{i}e_{k}, \\[2mm]
&\Sigma^a=\left[\frac{1}{2}(D_c\/v_b + D_b\/v_c)-\frac{1}{3}D_k v^k h_{cb}\right]N^{ac}\/e^b, \qquad
\Omega^a=\frac{1}{2}\varepsilon^{khij}N^a{}_h\/v_{k}\nabla_i\/v_{j}; 
\end{eqnarray}
\item tensor components
\begin{eqnarray}\label{kinematical_quantities tensors}
&\Sigma_{ab}=\frac{1}{2}\left(N^r{}_a\/N^k{}_{b}+N^r{}_{b}\/N^k{}_{a}-N_{ab}N^{kr}\right)\sigma_{rk}, \qquad
\zeta_{ab}=\frac{1}{2}\left(N^r{}_a\/N^k{}_{b}+N^r{}_{b}N^k{}_{a}-N_{ab}N^{kr}\right)\nabla_r\/e_{k},
\end{eqnarray}
\end{itemize}
where $\sigma_{rk}:=D_{\langle r }v_{k\rangle}$ is the shear tensor.

According to the $(1+1+2)$-splitting, the energy-momentum tensor of a given matter field can be decomposed in the form
\begin{eqnarray}\label{energytensor_generale}
\label{T}
&\!\!\!\!T_{ab}\!=\!\mu v_av_b\!-\!p(N_{ab}\!-\!e_ae_b)
\!+\!\frac{1}{2}\Pi(N_{ab}\!+\!2e_ae_b)
\!+\!(\Pi_{a}e_{b}\!+\!\Pi_{b}e_{a})\!+\!\Pi_{ab}
\!-\!Q(e_{a}v_{b}\!+\!e_{b}v_{a})\!+\!(Q_{a}v_{b}+Q_{b}v_{a}),
\end{eqnarray}
after having defined the thermodynamic quantities
\begin{subequations}\label{quantità_termodinamiche_generali}
\begin{align}
\label{energy_general}\mu &=T_{ab}v^av^b\\
\label{isopress_general}p &=-\frac{1}{3}T_{ab}(N^{ab}-e^a e^b)\\
\label{momentum_general}Q &=T_{ab}e^av^b\\
\label{anisopress_general}\Pi &=\frac{1}{3}T_{ab}(N^{ab}+2e^a e^b)\\
\label{anisopress_vector}\Pi_a &=-T_{cd}N^c{}_ae^d\\
\label{anisopress_tensor}\Pi_{ab} &=\left(N^c{}_{a}N^d{}_{b}-\frac{1}{2}N_{ab}N^{cd}\right)T_{cd}\\
\label{momentum_vector}Q_a &=T_{cd}N^c{}_a v^d.
\end{align}
\end{subequations}
More in detail, $\mu$ is the energy density, $p$ is the isotropic pressure, $Q$ is the scalar part of the momentum density, $\Pi$ is the scalar part of the anisotropic stress, $Q_a$ and $\Pi_a$ are respectively the vector parts of the momentum density and the anisotropic pressure, and $\Pi_{ab}$ are the components of the shearing pressure tensor.


\section{The Dirac Theory in Polar form}\label{Section3}
We briefly review the main features of the polar formalism for spinor fields \cite{Fabbri:2023dgv}. To this end, let $\gamma^{\mu}$ ($\mu=0,\ldots,3$) be a set of Clifford matrices, $\gamma^5:=i\gamma^{0}\gamma^{1}\gamma^{2}\gamma^{3}$ defining the parity-odd matrix. Given a tetrad field $e_\mu:=e^{a}_{\mu}\,\partial_a$, we denote by $\gamma^a :=e^a_{\mu}\gamma^{\mu}$. A spinor field $\psi$ is called regular if it satisfies either conditions $\bar\psi\psi \not=0$ or $\bar\psi\gamma^5\psi\not=0$, $\bar\psi:=\psi^{\dagger}\gamma^0$ being the adjoint spinor. Every regular spinor field $\psi$ can always be expressed (in chiral representation) as
\begin{eqnarray}
&\psi=\sqrt{\frac{\rho}{2}}e^{-\frac{i}{2}\beta\gamma^{5}}
\boldsymbol{L}^{-1}\left(\begin{tabular}{c}
$1$\\
$0$\\
$1$\\
$0$
\end{tabular}\right)
\label{spinor}
\end{eqnarray}
where the scalar $\rho$ and the pseudo-scalar $\beta$ are called respectively modulus and chiral angle, and $\boldsymbol{L}$ is a complex matrix with the structure of a complex Lorentz transformation \cite{jl1}. The polar form \eqref{spinor} allows us to express the bilinear quantities associated with a spinor field $\psi$ in the form
\begin{eqnarray}\label{bilinears}
&i\bar{\psi}\gamma^{5}\psi=\rho\sin{\beta},\label{sin}\ \ \ \ \ \ \ \ 
\bar{\psi}\psi=\rho\cos{\beta},\label{cos}\\
&\bar{\psi}{\gamma}^{a}\gamma^{5}\psi=\rho s^{a},\label{S}\ \ \ \ \ \ \ \ 
\bar{\psi}{\gamma}^{a}\psi=\rho u^{a},\label{U}
\end{eqnarray}
where the unit vector fields $u^a$ and $s^a$ satisfy the conditions
\begin{equation}\label{orthonormality}
u_{a}u^{a}=-s_{a}s^{a}=1 \quad {\rm and} \quad u_{a}s^{a}=0
\end{equation}
known as Fierz identities. Eqs. \eqref{orthonormality} reduce the degrees of freedom, represented by velocity $u^a$ and spin $s^a$ together, to five. Such degrees of freedom can be identified in the three space components of the velocity and the two angles that (in the rest frame) the spin forms with a given axis (in this paper, the third one).

At differential level, it is possible to show that there always exists a real vector $P_{a}$ and a tensor $R_{ija}=-R_{jia}$ in terms of which the covariant derivative of the spinor field in polar form can be written as
\begin{equation}\label{spinor_derivative}
\boldsymbol{\nabla}_{a}\psi=(-\frac{i}{2}\nabla_{a}\beta\gamma^{5}
+\frac{1}{2}\nabla_{a}\ln{\rho}\mathbb{I}
-iP_{a}\mathbb{I}-\frac{1}{4}R_{ija}\gamma^{i}\gamma^{j})\psi
\end{equation}
as discussed in \cite{Fabbri:2023dgv} and references therein. If there were no other term but $P_{a}$, one could write $i\boldsymbol{\nabla}_{a}\psi=P_{a}\psi$ in which $P_{a}$ would be seen as the momentum. The tensor $R_{ija}$ is called tensorial connection and it verifies the identities
\begin{eqnarray}\label{du}
\nabla_{a}s_{i}=R_{jia}s^{j},\ \ \ \ \ \ \ \ \ \ \ \ \ \ \ \ \nabla_{a}u_{i}=R_{jia}u^{j}
\end{eqnarray}
The latter can be inverted as
\begin{equation}\label{tensconn}
R_{abc}=u_{a}\nabla_{c}u_{b}-u_{b}\nabla_{c}u_{a}
+s_{b}\nabla_{c}s_{a}-s_{a}\nabla_{c}s_{b}
+(u_{a}s_{b}-u_{b}s_{a})\nabla_{c}u_{k}s^{k}
+2\varepsilon_{abij}u^{i}s^{j}V_{c}
\end{equation}
in terms of a vector $V^{c}$ for now not specified, but which is straightforward to interpret. In fact, plugging (\ref{tensconn}) back into (\ref{spinor_derivative}) one finds (see \cite{Fabbri:2023dgv} for a detailed proof)
\begin{equation}
\boldsymbol{\nabla}_{c}\psi=\left[-\frac{i}{2}\nabla_{c}\beta\gamma^{5}
+\frac{1}{2}\nabla_{c}\ln{\rho}\mathbb{I}
-i(P_{c}-V_{c})\mathbb{I}
-\frac{1}{4}u_{a}\nabla_{c}u_{b}[\gamma^{a},\gamma^{b}]
-\frac{1}{4}s_{b}\nabla_{c}s_{a}[\gamma^{a},\gamma^{b}]
-\frac{1}{4}u_{a}s_{b}\nabla_{c}u_{k}s^{k}[\gamma^{a},\gamma^{b}]\right]\psi.
\end{equation}
Because $V_{c}$ combines with the momentum and all remaining terms are derivatives of the spinor bilinears, this expression shows that the combination $P_{a}-V_{a}$ collects all information about the spinor derivatives that cannot be found in the derivatives of the spinor bilinears. This information, then, can only be found in the global phase, whose gradient is the momentum. Thus, the difference $P_{a}\!-\!V_{a}$ plays the role of effective momentum.

Using \eqref{spinor_derivative}, the polar form the Dirac equation $i\gamma^a\boldsymbol{\nabla}_{a}\psi -m\psi=0$ results to be equivalent to the pair
\begin{subequations}\label{Dirac_eq_polar}
\begin{align}
\label{b} &\nabla_{a}\beta+B_{a}-4P^{b}u_{[b}s_{a]}
+2ms_{a}\cos{\beta}=0, \\
\label{m} &\nabla_{a}\ln{\rho}+R_{a}-2P^{b}u^{c}s^{d}\varepsilon_{abcd}
+2ms_{a}\sin{\beta}=0,
\end{align}
\end{subequations}
where $R_{a}:= R_{ab}^{\phantom{ab}b}$ and $B_{a}:=\frac{1}{2}\varepsilon_{abcd}R^{bcd}$ (see \cite{Fabbri:2025ffi}).

Eventually, in polar form, the energy-momentum tensor of the Dirac field
\begin{equation}\label{Dirac_energy_tensor}
T^{ab}=
\frac{i}{8}\left(\bar{\psi}{\gamma}^{a}{\nabla}^{b}\psi
-{\nabla}^{b}\bar{\psi}{\gamma}^{a}\psi
+\bar{\psi}{\gamma}^{b}{\nabla}^{a}\psi
-{\nabla}^{a}\bar{\psi}{\gamma}^{b}\psi\right)
\end{equation}
can be expressed as
\begin{equation}\label{Dirac_energy_tensor_polar_form}
T^{ab}=
\frac{1}{4}\rho\left[P^{b}u^{a}+P^{a}u^{b}
+\frac{1}{2}\nabla^{a}\beta s^{b}
+\frac{1}{2}\nabla^{b}\beta s^{a}
-\frac{1}{4}R_{ij}^{\phantom{ij}a}\varepsilon^{bijk}s_{k}
-\frac{1}{4}R_{ij}^{\phantom{ij}b}\varepsilon^{aijk}s_{k}\right].
\end{equation}
\section{(1+1+2)-splitting of the polar formalism}\label{Section4}
In this section, we present the $(1+1+2)$ covariant decomposition of the polar formalism we briefly reviewed in the previous Section. In previous works \cite{Fabbri:2025ffi,VDFC}, such a decomposition was performed using the vectors $u^i$ and $s^i$ as generators of the congruences. Here we aim to generalize that treatment by considering instead a generic pair of timelike and spacelike vectors $(v^i,e^i)$ that are nevertheless still coplanar with respect to $(u^i,s^i)$. 

In view of the orthonormality conditions that the pairs $(v^i,e^i)$ and $(u^i,s^i)$ must meet separately, this allows us to relate the two pairs of vector fields via a single pseudo-scalar function $\eta$ as follows
\begin{subequations}\label{frame}
\begin{align}
&u^{i} = \cosh{\eta}\, v^{i} - \sinh{\eta}\, e^{i }\\
&s^{i} = -\sinh{\eta}\, v^{i} + \cosh{\eta}\, e^{i}
\end{align}
\end{subequations}
with inverse relations given by
\begin{subequations}\label{frameinv}
\begin{align}
&v^{i} = \cosh{\eta}\, u^{i} + \sinh{\eta}\, s^{i} \\
&e^{i} = \sinh{\eta}\, u^{i} + \cosh{\eta}\, s^{i}.
\end{align}
\end{subequations}
From \eqref{frame} and \eqref{frameinv} we have
\begin{equation}
u^{j}s^{k}\varepsilon_{jkab} = v^{j}e^{k}\varepsilon_{jkab}
\end{equation} 
or equivalently
\begin{equation}
s^{a}s^{b}-u^{a}u^{b}=e^{a}e^{b}-v^{a}v^{b}.
\end{equation} 
In particular, due to relations \eqref{frame}, the covariant derivatives of the vector fields $u^i$ and $s^i$ can be expressed in terms of the covariant derivatives of $\eta$, $v^i$ and $e^i$ as
\begin{subequations}\label{dercov_us}
\begin{align}
&\nabla_{a}u_{j} = -\nabla_{a}\eta\, s_{j} + \cosh{\eta}\, \nabla_{a}v_{j} - \sinh{\eta}\, \nabla_{a}e_{j}\\
&\nabla_{a}s_{j} = -\nabla_{a}\eta\, u_{j} - \sinh{\eta}\, \nabla_{a}v_{j} + \cosh{\eta}\, \nabla_{a}e_{j}
\end{align}
\end{subequations}
inserting \eqref{dercov_us} into \eqref{tensconn}, the tensorial connection $R_{abc}$ can be written in the form
\begin{eqnarray}
R_{abc} &=& v_{a}\nabla_{c}v_{b}-v_{b}\nabla_{c}v_{a}
+ e_{b}\nabla_{c}e_{a}-e_{a}\nabla_{c}e_{b}
+(v_{a}e_{b}-v_{b}e_{a})\nabla_{c}v_{k}e^{k}
- (v_{a}e_{b}-v_{b}e_{a})\nabla_{c}\eta + 2\varepsilon_{ab}V_{c}
\end{eqnarray}
and using \eqref{covariant_derivatives_un} we can further write
\begin{eqnarray}\label{Rfullev}
\nonumber
R_{abc} &=& v_{a}\Sigma_{c b}-v_{b}\Sigma_{c a} + e_{b}\zeta_{c a}-e_{a}\zeta_{c b} - (e_{a}\alpha_{b}-e_{b}\alpha_{a})v_{c} + (v_{a}\mathcal{A}_{b}-v_{b}\mathcal{A}_{a})v_{c} \\
\nonumber
&-& (v_{a}\Sigma_{b}-v_{b}\Sigma_{a})e_{c} + (e_{a}a_{b}-e_{b}a_{a})e_{c} + (v_{a}\varepsilon_{bk}\Omega^{k} - v_{b}\varepsilon_{ak}\Omega^{k}) e_{c} \\
\nonumber
&+& \frac{1}{3}\Theta(v_{a}h_{c b}-v_{b}h_{c a}) + \frac{1}{2}\Sigma(v_{a}N_{c b}-v_{b}N_{c a}) - \frac{1}{2}\phi(e_{a}N_{c b}-e_{b}N_{c a}) \\
\nonumber
&+& \Omega(v_{a}\varepsilon_{c b}-v_{b}\varepsilon_{c a}) - \xi(e_{a}\varepsilon_{c b}-e_{b}\varepsilon_{c a}) \\
&-& (v_{a}e_{b}-v_{b}e_{a})(Av_{c}-\Sigma e_{c}+\Sigma_{c}+\varepsilon_{ck}\Omega^{k}+\nabla_{c}\eta) + 2\varepsilon_{ab}V_{c}
\end{eqnarray}
in terms of the kinematical quantities related to the time-like and the space-like congruences.

By replacing \eqref{Rfullev} into \eqref{Dirac_eq_polar} and projecting, we can decompose the Dirac equations into the set of equations
\begin{subequations}\label{Dirac_decomposed_general}
\begin{align}
\label{direceq1}&\dot{\ln{\rho}} - \hat{\eta} + \Theta - 2m\sinh{\eta}\sin{\beta} = 0\\
\label{diraceq2}&\hat{\ln{\rho}} - \dot{\eta} + \phi - A - 2m\cosh{\eta}\sin{\beta} = 0\\
\label{direceq3}&\delta_{i}\beta - \varepsilon_{ik}(\alpha^{k}+2\varepsilon^{kj}\Omega_{j}+\nabla^{k}\eta) = 0 \\
\label{diracmomentum1}&2(P-V)_i v^i = 2m\cosh{\eta}\cos{\beta} - 2\Omega - \hat\beta\\
\label{diracmomentum2}&2(P-V)_i e^i = 2m\sinh{\eta}\cos{\beta} - 2\xi - \dot\beta\\
\label{diracmomentum3}&2(P-V)_{i}\varepsilon^{ik} = -N^{ik}\nabla_{i}\ln{\rho} + \mathcal{A}^{k} - a^{k}.
\end{align}
\end{subequations}
In a similar way, the energy-momentum tensor \eqref{Dirac_energy_tensor_polar_form} can be recast in the form \eqref{energytensor_generale} with associated thermodynamic quantities given now by
\begin{subequations}\label{energy0}
\begin{align}
\label{energy01}\mu &= \frac{1}{4}\rho\left[2(P-V)_i v^i \cosh{\eta} - \sinh{\eta} v^i\nabla _i \beta \right] \\
\label{enrgy02}p &= \frac{1}{12}\rho\left[2(P-V)_i e^i \sinh{\eta} - 2\Omega \cosh{\eta} + 2\xi \sinh{\eta} - \cosh{\eta} e^i \nabla_i\beta \right] \\
\label{energy03}Q &= \frac{1}{8}\rho\left[2(P-V)_i(e^i \cosh{\eta} + v^i \sinh{\eta}) - (\cosh{\eta}v^i + \sinh{\eta}e^i) \nabla_i \beta \right]\\
\label{anpress0}\Pi &= \frac{1}{6}\rho\left[2(P-V)_i e^i \sinh{\eta} + \Omega \cosh{\eta} - \xi \sinh{\eta} - \cosh{\eta} e^i \nabla_i \beta \right] \\
\label{energy05}Q^{a} &= \frac{1}{8}\rho\left[2(P-V)_{i}N^{ia}\cosh{\eta} - \sinh{\eta}N^{ia}\nabla_{i}\beta - \mathcal{A}_{i}\varepsilon^{ia}\cosh{\eta} + \alpha_{i}\varepsilon^{ia}\sinh{\eta}\right]\\
\label{energy06}\Pi^{a} &= -\frac{1}{8}\rho\left[2(P-V)_{i}N^{ia}\sinh{\eta} - \cosh{\eta}N^{ia}\nabla_{i}\beta + a_{i}\varepsilon^{ia}\sinh{\eta} - (\Sigma_{i}\varepsilon^{ia}-\Omega^{a})\cosh{\eta}\right]\\
\label{energy07}\Pi^{ab} &= \frac{1}{8}\rho\left[(\varepsilon^{a}_{\phantom{a}j}\Sigma^{bj} + \varepsilon^{b}_{\phantom{b}j}\Sigma^{aj})\cosh{\eta} - (\varepsilon^{a}_{\phantom{a}j}\zeta^{bj} + \varepsilon^{b}_{\phantom{b}j}\zeta^{aj})\sinh{\eta}\right].
\end{align}
\end{subequations}
A few remarks are needed: the first is that eqs. \eqref{energy0} (together with eq. \eqref{energytensor_generale}) provide us with a complete and consistent description of the energy-momentum tensor in terms of $\eta$, $\rho$, $\beta$, $P_i - V_i$ and the kinematical quantities associated with the time-like and space-like congruences. The tensorial connection $R_{ija}$ can be tied through eq. (\ref{du}) to velocity and spin. Instead, there is no equivalent of eq. (\ref{du}) for $P_{i}$ and $V_{i}$, which, therefore, cannot be determined in terms of the fundamental vector fields of the underlying geometry. As such, they must be treated as external and unknown fields. However, in the quantities \eqref{energy0} only the difference $P_i-V_i$ actually appears. The latter is entirely determined by the last  Dirac equations \eqref{diracmomentum1}-\eqref{diracmomentum3}. The idea is then to use the eqs. \eqref{diracmomentum1}-\eqref{diracmomentum3} to determine the vector field $P_i-V_i$. The information lost in doing this is restored by the conservation laws which yield exactly four scalar equations.
\section{Chiral scalings}\label{CsApp}
In this section, we discuss the relations \eqref{frameinv} from the perspective of the original spinorial components. To be more specific, let us write eqs. \eqref{frameinv} after multiplying by $\rho$, getting
\begin{gather}
\rho v^{i} = \cosh{\eta}\,\bar{\psi}{\gamma}^{i}\psi
+ \sinh{\eta}\,\bar{\psi}{\gamma}^{i}\gamma^5\psi\label{A1}\\
\rho e^{i} = \sinh{\eta}\,\bar{\psi}{\gamma}^{i}\psi
+ \cosh{\eta}\,\bar{\psi}{\gamma}^{i}\gamma^5\psi.\label{A2}
\end{gather}
Clearly, one could ask whether it is possible to have both expressions \eqref{A1} and \eqref{A2} induced by an assigned transformation acting on the spinor field. The answer is positive, although such a transformation is not (induced by) a Lorentz one.

Indeed, the transformation in question has the structure
\begin{gather}
\psi \to \psi'\!:=\!e^{\eta\gamma^5/2}\psi
\!=\![\cosh{(\eta/2)}\mathbb{I}\!+\!\sinh{(\eta/2)}\gamma^5]\psi,\label{chirscal}
\end{gather}
also giving
\begin{gather}
\bar{\psi}'\!=\!\bar{\psi}e^{-\eta\gamma^5/2}.
\end{gather}
Therefore we have
\begin{gather}\label{eq58}
\left(\bar{\psi}{\gamma}^{i}\psi\right)'
=\bar{\psi}e^{-\eta\gamma^5/2}{\gamma}^{i}e^{\eta\gamma^5/2}\psi
=\bar{\psi}{\gamma}^{i}e^{\eta\gamma^5/2}e^{\eta\gamma^5/2}\psi
=\bar{\psi}{\gamma}^{i}e^{\eta\gamma^5}\psi
=\bar{\psi}{\gamma}^{i}[\cosh{\eta}\mathbb{I}+\sinh{\eta}\gamma^5]\psi=\\
\nonumber
=\cosh{\eta}\bar{\psi}{\gamma}^{i}\psi
+\sinh{\eta}\bar{\psi}{\gamma}^{i}\gamma^5\psi.
\end{gather}
In polar notation, defining $U^i:=\bar{\psi}{\gamma}^{i}\psi=\rho u^i$ and $S^i:=\bar{\psi}{\gamma}^{i}\gamma^5\psi=\rho s^i$, equation \eqref{eq58} assumes the simpler form
\begin{gather}
U_{i}'=\cosh{\eta}U_{i}+\sinh{\eta}S_{i}.
\end{gather}
It is a straightforward matter to see that the relation
\begin{gather}
S_{i}'=\cosh{\eta}S_{i}+\sinh{\eta}U_{i}
\end{gather}
also holds, so that both eqs. (\ref{A1}) and (\ref{A2}) are in fact induced by transformation (\ref{chirscal}) (provided that $v_{i}\!=\!u_{i}'$ and $e_{i}\!=\!s_{i}'$). Writing eqs. (\ref{A1}) and (\ref{A2}) in matrix form gives
\begin{gather}
\left(\begin{array}{c}
\rho v^{i}\\
\rho e^{i}
\end{array}\right)=\left(\begin{array}{cc}
\cosh{\eta} & \sinh{\eta}\\
\sinh{\eta} & \cosh{\eta}
\end{array}\right)\left(\begin{array}{c}
U^{i}\\
S^{i}
\end{array}\right),
\end{gather}
which might suggest that the transformation is actually a boost. However, this is not the case because, unlike the above transformation, a boost does not mix velocity and spin components. As a matter of fact, a quick use of the polar form (\ref{spinor}) would reveal that (\ref{chirscal}) acts as if it was a conformal scaling, but one for which the two chiral components are treated in opposte ways (specifically, positive values of $\eta$ would shrink the left part down and inflate the right part up, and vice versa).

\section{Locally Rotationally Symmetric space-times}\label{Section2bis}
In this work, we focus on Locally Rotationally Symmetric (LRS) space-times. In these geometries, at every point of space-time, the vector field $e^i$ identifies a local axis of symmetry. All observations are identical under rotations around $e^i$. In other words, observations are the same in all spatial directions perpendicular to $e^i$. As a consequence, all tensors representing physical quantities must have null projections into the two-spaces orthogonal to both $v^i$ and $e^i$. According to this geometrical setting, the covariant derivatives \eqref{covariant_derivatives_un} reduce to
\begin{subequations}\label{LRS_derivatives}
\begin{align}
\nabla_i v_j &= \Sigma\!\left(e_i e_j + \tfrac{1}{2}N_{ij}\right)
 + \tfrac{1}{3}\Theta\!\left(N_{ij} - e_i e_j\right)
 - A v_i e_j + \Omega \varepsilon_{ij} \label{lrs_1},\\
\nabla_i e_j &= \tfrac{1}{2}\phi N_{ij} + \xi \varepsilon_{ij}
 - A v_i v_j + \left(\Sigma - \tfrac{1}{3}\Theta\right)e_i v_j \label{lrs_2}.
\end{align}
\end{subequations}
Moreover, the following identities necessarily hold:
\begin{subequations}\label{LRS_identities}
\begin{align}
\dot{v}^a &= -A e^a, & \omega^a &:=\frac{1}{2}\varepsilon^{abc}D_bv_c =-\Omega e^a, \\
\sigma_{ab} &:=D_{\langle r }v_{k\rangle} = \Sigma\!\left(e_a e_b + \tfrac{1}{2}N_{ab}\right), &
E_{ab} &= E\!\left(e_a e_b + \tfrac{1}{2}N_{ab}\right), &
H_{ab} &= H\!\left(e_a e_b + \tfrac{1}{2}N_{ab}\right),
\end{align}
\end{subequations}
where $\dot{v}^a$ is the four-acceleration, $\omega^a$ is the vorticity vector field, $\sigma_{ab}$ is the shear tensor, $E_{ab}$ and $H_{ab}$ denote the electric and the magnetic part of the Weyl tensor.

Still in accordance with the LRS geometry, the two-spatial quantities \eqref{anisopress_vector}, \eqref{anisopress_tensor}, and \eqref{momentum_vector} must be set equal to zero. In such a circumstance, the energy--momentum tensor \eqref{energytensor_generale} reduces to
\be\label{EMTLRS}
T_{ab}\!=\!\mu v_{a}v_{b}\!-\!p(N_{ab}\!-\!e_{a}e_{b})\!-\!Q(e_{a}v_{b}\!+\!e_{b}v_{a})\!+\!\frac{1}{2}\Pi(N_{ab}\!+\!2e_{a}e_{b}),
\ee  
where the quantities $\mu$, $p$, $Q$ and $\Pi$ are given by equations \eqref{energy_general}-\eqref{anisopress_general}. For practical purposes, it is often useful to write the tensor \eqref{EMTLRS} as
\be\label{EMTLRS2}
T_{ab}\!=\!\mu v_{a}v_{b}\!+\!p_re_{a}e_{b}\!-p_oN_{ab}-\!Q(e_{a}v_{b}\!+\!e_{b}v_{a})\!,
\ee  
where
\be\label{radortpress}
p_r := p + \Pi  \qquad {\rm and} \qquad 
p_o := p - \frac{1}{2} \Pi .
\ee
$p_r$ is the pressure along the preferred spatial direction defined by the vector field $e^a$ (tangent to the space-like congruence), whereas $p_o$ describes the pressure in the orthogonal subspace.

A LRS space-time filled by a given matter field is then completely characterized by the following set of scalar quantities 
\begin{equation}\label{LRS_scalars}
\{A,\Theta,\Sigma,\Omega,\phi,\xi,E,H,\mu,p,Q,\Pi\}.
\end{equation}
In the signature $(+,-,-,-)$, the covariant equations for the variables \eqref{LRS_scalars} were derived in \cite{VDFC} and they are expressed as 
\\
\\
{\it Evolution equations}:
\begin{subequations}\label{evolutions_equations}
\begin{align}
\label{dot_Omega} \dot{\Omega} &= -A\xi-\frac{2}{3}\Theta\Omega-\Omega\Sigma\\
\label{dot_H} \dot{H} &= 3E\xi+\frac{3}{2}\Pi\xi-\Theta H -\frac{3}{2}H\Sigma -\Omega Q \\
\label{dot_phi} \dot{\phi} &= - \left(\Sigma+\frac{2}{3}\Theta\right)\left(A+\frac{1}{2}\phi\right) + 2\Omega\xi - Q \\
\label{dot_xi} \dot{\xi} &= -\left(\frac{1}{2}\Sigma+\frac{1}{3}\Theta\right)\xi - \Omega\left(\frac{1}{2}\phi+A\right) + \frac{1}{2}H .
\end{align}
\end{subequations}
\\
{\it Propagation equations}:
\begin{subequations}\label{propagation_equations}
\begin{align}
\label{hat_Omega}\hat{\Omega} &= - \Omega\left(A+\phi\right)\\
\label{hat_phi} \hat{\phi} &= -\frac{1}{2}\phi^{2} + 2\xi^{2} - \left(\Sigma-\frac{1}{3}\Theta\right)\left(\Sigma+\frac{2}{3}\Theta\right) - \frac{2}{3}\mu -\frac{1}{2}\Pi + E \\
\label{hat_xi} \hat{\xi} &= - \left(\Sigma-\frac{1}{3}\Theta\right)\Omega - \phi\xi \\ 
\label{hat_Theta_Sigma} \frac{2}{3}\hat{\Theta} + \hat{\Sigma} &= -\frac{3}{2}\Sigma\phi + 2\Omega\xi - Q \\
\label{hat_E} \hat{E}-\frac{1}{2}\hat{\Pi}+\frac{1}{3}\hat{\mu} &= - \frac{3}{2}\phi\left(E-\frac{1}{2}\Pi\right) - 3\Omega H - Q\left(\frac{1}{3}\Theta+\frac{1}{2}\Sigma\right)\\
\label{hat_H} \hat{H} &= - \frac{3}{2}\phi H - \Omega\left(-3E+\mu+p-\frac{1}{2}\Pi\right) + Q\xi.
\end{align}
\end{subequations}
\\
{\it Evolution--Propagation equations}:
\begin{subequations}\label{evolution_propagation_equations}
\begin{align}
\label{Raychaudhuri1+1+2}  
\dot{\Theta} + \hat{A} &= - A\phi + A^{2} - \frac{1}{3}\Theta^{2} - \frac{3}{2}\Sigma^{2} + 2\Omega^{2} - \frac{1}{2}\left(\mu+3p\right) \\
\label{dot_mu} \dot{\mu} - \hat{Q} &= - \left(\mu + p\right)\Theta + Q\phi - 2AQ +\frac{3}{2}\Sigma\Pi \\
\label{dot_Q} \dot{Q}-\hat{\Pi} -\hat{p} &= + \frac{3}{2}\Pi\phi - A\left(\mu + p + \Pi\right) - \frac{4}{3}Q\Theta + Q\Sigma \\
\label{dot_Sigma} \dot{\Sigma}-\frac{2}{3}\hat{A} &= - \frac{1}{3}A\phi + \frac{1}{2}\Sigma^{2} + \frac{2}{3}\Omega^{2} - \frac{2}{3}A^{2} - \frac{2}{3}\Theta\Sigma - E - \frac{1}{2}\Pi \\
\label{dot_E} \dot{E} - \frac{1}{2}\dot{\Pi} + \frac{1}{3}\hat{Q} &= -E\Theta - \frac{3}{2}E\Sigma - 3H\xi + \frac{2}{3}AQ + \frac{1}{6}Q\phi - \frac{1}{2}\left(\mu+p\right)\Sigma + \frac{1}{6}\Theta\Pi - \frac{1}{4}\Pi\Sigma .
\end{align}
\end{subequations}
\\
{\it Constraint equation}:
\begin{equation}\label{constraint_H}
H = 2A\Omega + \Omega\phi - 3\xi\Sigma.
\end{equation}
The analysis of the consistency and integrability conditions for equations \eqref{evolutions_equations}-\eqref{constraint_H} was carried out in \cite{VDFC}. In particular, for LRS space-times of types I, II, and III, such consistency and integrability conditions are given by the two constraints
\begin{subequations}\label{condizioni_integrabilità}
\begin{align}
&\label{rif1} 
\phi\xi=\left(\Sigma+\frac{2}{3}\Theta\right)\Omega, \\[1mm]
& \label{rif2}
\left(p+\mu+\Pi\right)\xi\Omega=Q\left(\Omega^{2}+\xi^{2}\right) .
\end{align}
\end{subequations}
In deriving conditions \eqref{condizioni_integrabilità}, a crucial role is played by the kinematic relations
\begin{subequations}
\begin{align}
&\label{consistency_f}
\dot{f}\Omega=\hat{f}\xi,\\[1mm]
&\label{inteffe} 
\hat{\dot{g}} - \dot{\hat{g}}= + A\dot{g} - \Sigma\hat{g} + \frac{1}{3}\Theta\hat{g},
\end{align}
\end{subequations}
that every covariantly defined scalar quantity $f$ ($\delta_{a}f=0$) and every scalar function $g$ must verify. For further details, see \cite{VDFC}.

Originally, LRS space-times were classified for a perfect fluid ($Q=\Pi=0$). In particular, the requirement that the momentum density vanishes in \eqref{rif2} implies $\Omega\xi=0$. From this condition, three distinct classes arise:
\begin{itemize}
\item LRSI space-times: $\Omega\neq0$ and $\xi=0$;
\item LRSII space-times: $\Omega=\xi=0$;
\item LRSIII space-times: $\Omega=0$ and $\xi\neq0$.
\end{itemize}
For a generic fluid, the condition \eqref{rif2} always implies $Q=0$ in LRS space-times of types I and III, whereas in LRSII space-times the momentum density $Q$ can be, in general, non-zero.
\section{Spinorial fluid in LRS space-times}\label{Section5}
In this section, we implement the matching between the covariant $(1+1+2)$ approach and the polar formalism, which has been presented in the previous Sections. Focusing exclusively on LRS space-times of types I, II and III, the proposed geometrical construction generalizes the approach given in \cite{VDFC}, where the unit vector fields $u^i$ and $s^i$ had been chosen to coincide with the generators $v^i$ and $e^i$ of the temporal and spatial congruences, respectively. Here we weaken that prescription by requiring that the pairs of vector fields $(v^i,e^i)$ and $(u^i,s^i)$ be coplanar, as in Section \ref{Section4}.

According to the requirements of LRS space-times, all two-spatial geometric quantities must be zero. Therefore, the energy-momentum tensor assumes the simplified form \eqref{EMTLRS}, where the thermodynamic quantities $\mu$, $p$, $Q$ and $\Pi$ come from equations (\ref{energy01})-(\ref{anpress0}), after omitting all the two-spatial components. In detail, by making use of equations \eqref{diracmomentum1} and \eqref{diracmomentum2} as well as of the notations $\dot{f}:=v^i\nabla_i\/f$ and $\hat{f}:=e^i\nabla_i\/f$ for every scalar function $f$, we have:
\begin{subequations}\label{quantità_termodinamiche}
\begin{align}
\mu&=\frac{1}{2}\rho\left[\left(m\cosh{\eta}\cos{\beta}-\Omega
-\frac{\hat{\beta}}{2}\right)\cosh{\eta}-\frac{1}{2}\dot{\beta}\sinh{\eta}\right],\label{energy}\\
p&=-\frac{1}{12}\rho\left[\left(2\Omega+\hat{\beta}\right)\cosh{\eta}+\left(\dot{\beta}-2m\sinh{\eta}\cos{\beta}\right)\sinh{\eta}\right],\label{pressure}\\
Q&=-\frac{1}{4}\rho\left[\left(\dot{\beta}+\xi\right)\cosh{\eta}+\left(\hat{\beta}+\Omega-2m\cosh{\eta}\cos{\beta}\right)\sinh{\eta}\right],\label{momentum}\\
\Pi&=-\frac{1}{6}\rho\left[\left(\hat{\beta}-\Omega\right)\cosh{\eta}+\left(\dot{\beta}+3\xi-2m\sinh{\eta}\cos{\beta}\right)\sinh{\eta}\right].\label{anpressure}
\end{align}
\end{subequations}
The remaining Dirac equations are expressed as 
\begin{subequations}\label{Dirac_equations}
\begin{align}
\label{dotrho}&\dot{\ln{\rho}}-\hat{\eta}+\Theta-2m\sinh{\eta}\sin{\beta}=0,\\
\label{hatrho}&\hat{\ln{\rho}}-\dot{\eta}+\phi-A-2m\cosh{\eta}\sin{\beta}=0.
\end{align}
\end{subequations}
We will discuss the covariant equations for a self-gravitating Dirac field in LRS space-times of types I, II, and III. In general, we will consider the case of non perfect spinorial fluid, dealing with the perfect case as a particular one. In LRS space-times of classes I and III, the momentum density $Q$ vanishes. Hence, from eq. \eqref{momentum} we have
\begin{equation}\label{perfectfluidconstr1}
\left(\dot{\beta}+\xi\right)\cosh{\eta}+\left(\hat{\beta}+\Omega-2m\cosh{\eta}\cos{\beta}\right)\sinh{\eta}=0.
\end{equation}
In these classes, the difference between a perfect and a non perfect fluid is due to the presence or absence of anisotropic pressure $\Pi$. In LRSII space-times, the momentum density $Q$ can be non-zero; in that case, the restriction to the perfect fluid case requires to impose that both $Q$ and $\Pi$, given by eqs. \eqref{momentum} and \eqref{anpressure}, be zero. The vanishing of the anisotropic pressure \eqref{anpressure} yields the equation
\begin{equation}
\label{perfectfluidconstr2}\left(\hat{\beta}-\Omega\right)\cosh{\eta}+\left(\dot{\beta}+3\xi-2m\sinh{\eta}\cos{\beta}\right)\sinh{\eta}=0.
\end{equation}
Eqs. \eqref{perfectfluidconstr2} and \eqref{perfectfluidconstr1} can be solved for $\dot{\beta}$ and $\hat{\beta}$ as
\begin{subequations}\label{dothatbeta}
\begin{align}
\label{dothatbeta1}&\dot{\beta}=-2\xi+\xi\cosh{2\eta}+2m\cos{\beta}\sinh{\eta}-\Omega\sinh{2\eta},\\
\label{dothatbeta2}&\hat{\beta}=\Omega\cosh{2\eta}-\xi\sinh{2\eta}.
\end{align}
\end{subequations}
The kinematical constraint $\dot{\beta} \, \Omega = \hat{\beta} \, \xi$ (see eq. \eqref{consistency_f}) then yields the relation
\begin{eqnarray}\label{betaconstraint}
\left[m\Omega\cos{\beta}+\left(\xi^2-\Omega^2\right)\cosh{\eta}\right]\sinh{\eta}=\xi\Omega=0,
\end{eqnarray}
which has to be satisfied when the spinorial fluid is perfect.
\subsection{LRSI space-times}
LRSI space-times are characterized by the condition $\xi=0$, $\Omega \not =0$ and $Q=0$.
The kinematical constraint \eqref{consistency_f} implies $\dot{f}=0$ for every covariantly defined scalar $f$. From eqs. \eqref{dot_Omega} and  \eqref{hat_xi}, we deduce that $\Sigma=0$ and $\Theta=0$. 
After that, the covariant equations \eqref{dot_Omega}, \eqref{dot_H}, \eqref{dot_phi}, \eqref{hat_Theta_Sigma}, \eqref{dot_mu} and \eqref{dot_E} are automatically satisfied. The evolution equation \eqref{dot_xi} becomes identical to the constraint \eqref{constraint_H},
giving us the expression for $H$
\begin{equation}\label{H_bis}
H=2A\Omega + \phi\Omega.
\end{equation}
Also, by combining equations \eqref{Raychaudhuri1+1+2} and \eqref{dot_Sigma}, we obtain the explicit expression for E
\bea\label{eexpr} 
E=-A\phi+2\Omega^2-\frac{1}{3}\left(\mu+3p\right)-\frac{1}{2}\Pi .
\eea
Expressions \eqref{H_bis} and \eqref{eexpr} automatically verify the corresponding propagation equations \eqref{hat_H} and \eqref{hat_E}. This is seen through a direct check. Discarding the solution $\eta=0$ (already studied in \cite{VDFC}), from the vanishing of $Q$ we get the propagation equation for $\beta$ (see eq.  \eqref{perfectfluidconstr1})
\be\label{hatbeta_LRSInp}
\hat{\beta}=2m\cosh{\eta}\cos{\beta}-\Omega.
\ee
By inserting eq. \eqref{hatbeta_LRSInp} into eq. \eqref{quantità_termodinamiche}, we derive the representation of the thermodynamic quantities 
\be\label{quantità_termodinamiche_LRSI_np}
\begin{cases}
\mu=-\frac{1}{4}\rho\Omega\cosh{\eta}\\
p=-\frac{1}{12}\rho\left(2m\cos{\beta}+\Omega\cosh{\eta}\right)\\
\Pi=\frac{1}{6}\rho\left(2\Omega\cosh{\eta}-2m\cos{\beta}\right)
\end{cases}
\ee
which correspond to the equation of state
\begin{equation}\label{eosLRSInp}
\mu=p-\frac12\Pi = p_o.
\end{equation}
Eventually, making use of eqs. \eqref{hatbeta_LRSInp} and \eqref{quantità_termodinamiche_LRSI_np}, it is easily seen that the covariant equation \eqref{dot_Q} is identically verified.

To conclude, we end up with the final system of differential equations
 \bea\label{eq_finali_LRSI_non_perfect}
\begin{cases}
\hat{A}=-A\phi+A^2+2\Omega^2+\frac{1}{4}m\rho\cos{\beta}+\frac{1}{4}\Omega\rho\cosh{\eta} \\
\hat{\Omega}=-\Omega\left(A+\phi\right) \\
\hat{\phi}= \!-\frac{1}{2}\phi^{2}-A\phi+2\Omega^2+\frac{1}{2}m\rho\cos{\beta} \\
\hat{\rho}=\rho\left(2m\cosh{\eta}\sin{\beta}+A-\phi\right)\\
\hat{\beta}=2m\cosh{\eta}\cos{\beta}-\Omega\\
\hat{\eta}=- 2m\sinh{\eta}\sin{\beta}
\end{cases}
\eea
with unknowns $\{A,\Omega,\phi,\rho,\beta, \eta\}$. According to the Cauchy theorem, the dynamical system \eqref{eq_finali_LRSI_non_perfect} is well-posed. Assigned initial (boundary) data for all the unknowns on a given time-like hypersurface orthogonal to $e^i$, at least locally there exists a unique solution. A plot of a numerical solution of eqs. \eqref{eq_finali_LRSI_non_perfect} is shown in Section \ref{sezione_8}. 

\subsubsection*{\bf Perfect fluid case}

In such a circumstance, since $\Pi=0$, the expression \eqref{eexpr} reduces to
\bea\label{evin} 
E=-A\phi+2\Omega^2-\frac{1}{3}\left(\mu+3p\right) .
\eea
Again, expressions \eqref{H_bis} and \eqref{evin} make equations \eqref{hat_E} and \eqref{hat_H} automatically verified, as well as equations \eqref{dot_H}, \eqref{dot_phi}, \eqref{hat_Theta_Sigma} \eqref{dot_mu} and \eqref{dot_E} are identically satisfied. The remaining covariant and Dirac equations are
\begin{subequations}\label{eq_cov_LRSI_perfect}
\begin{align}
& \label{eq_cov_LRSI_perfect1} A\phi +\hat{A}- A^{2}-2\Omega^{2}+\frac{1}{2}\left(\mu+3p\right)=0 \\
& \label{eq_cov_LRSI_perfect3}\hat{p} - A\left(\mu+ p\right)=0 \\
& \label{eq_cov_LRSI_perfect7} \hat{\Omega}+\Omega\left(A+\phi\right)=0  \\
& \label{eq_cov_LRSI_perfect8} \hat{\phi}+\frac{1}{2}\phi^{2}+\frac{2}{3}\mu-E=0  \\
& \label{eq_cov_LRSI_perfect12} \hat{\ln{\rho}}- 2m\sin{\beta}\cosh{\eta} - A + \phi=0\\
& \label{eq_cov_LRSI_perfect13} \hat{\eta} + 2m\sinh{\eta}\sin{\beta}=0,
\end{align}
\end{subequations}
where $\mu$ and $p$ are given by equations \eqref{energy} and \eqref{pressure}. Moreover, eq. \eqref{dothatbeta2} now reads as
\begin{eqnarray}
\label{hat_beta_LRSI}\hat{\beta}=\Omega\cosh{2\eta}.
\end{eqnarray}
From eq. \eqref{dothatbeta1} (or, equivalently from the expression for $\Pi$ in \eqref{quantità_termodinamiche_LRSI_np}), we obtain the identity 
\begin{equation}\label{omega_LRSI}
\Omega=\frac{m\cos{\beta}}{\cosh{\eta}},
\end{equation}
which satisfies the constraint \eqref{betaconstraint}. Making use of equations \eqref{hat_beta_LRSI} and \eqref{omega_LRSI}, it is easily seen that the energy density and the pressure assume the simpler form
\begin{equation}\label{energy_pressure__LRSI}
\mu=-\frac{1}{4}m\rho\cos{\beta} \quad {\rm and} \quad p=-\frac{1}{4}m\rho\cos{\beta}.
\end{equation}
From equations \eqref{energy_pressure__LRSI} we read off the stiff equation of state 
\begin{equation}\label{eos_LRSIp}
p=\mu.
\end{equation}
The expression \eqref{omega_LRSI} for $\Omega$ must satisfy the corresponding propagation equation. By inserting eq. \eqref{omega_LRSI} in eq. \eqref{eq_cov_LRSI_perfect7}, we get the equation
\be\label{equazione_hatA_LRSI}
\frac{m\left[\left(A+\phi\right)\cosh{\eta}-m\sin{\beta}\right]\cos{\beta}}{\cosh^2{\eta}} =0,
\ee
which is satisfied for
\be\label{A_LRSI}
A=m\frac{sin\beta}{\cosh\eta} -\phi \quad \cup \quad \cos\beta = 0.
\ee
Due to eq. \eqref{omega_LRSI}, the solution $\cos{\beta} = 0$ implies $\Omega = 0$ which contradicts the LRSI assumption. Therefore, we focus only on the first of the solutions \eqref{A_LRSI}. In this regard, a direct check shows that equations \eqref{eq_cov_LRSI_perfect12}, \eqref{hat_beta_LRSI}, \eqref{energy_pressure__LRSI} and \eqref{A_LRSI} make the propagation equation \eqref{eq_cov_LRSI_perfect3} automatically satisfied. Instead, by inserting the expression \eqref{A_LRSI} for $A$ into eq. \eqref{eq_cov_LRSI_perfect1} and making use of all the previously obtained identities, we get the further constraint equation
\be\label{A_constr_LRSI}
\left(4m^2-5\phi^2-2m\rho\cos{\beta}\right)\cosh^2{\eta}+8m\phi\sin{\beta}\cosh{\eta}-4m^2\cos^2{\beta}-6m^2=0.
\ee
The latter can be easily solved for $\rho$ as
\be\label{rho_LRSI}
\rho=\frac{4m^2-5\phi^2}{2m\cos{\beta}}+\frac{4\phi\sin{\beta}}{\cos{\beta}\cosh{\eta}}-\frac{m\left(2cos^2{\beta}+3\right)}{\cos{\beta}\cosh^2{\eta}}.
\ee
Eventually, from equations \eqref{eq_cov_LRSI_perfect8}, \eqref{eq_cov_LRSI_perfect13} and \eqref{hat_beta_LRSI}, we derive the final set of differential equations
\begin{equation}\label{equazioni_finali_LRSI}
\begin{cases}
\hat\phi &=\dfrac{4m\phi\sin\beta\cosh{\eta} + 4m^2\cosh^2{\eta} - 3\phi^2\cosh^2{\eta} + 4m^2\cos^2{\beta} - 6m^2}{4\cosh^2{\eta}}\\
\hat\eta &=-2m\sinh\eta\sin\beta\\
\hat\beta &=\dfrac{m\cos\beta}{\cosh{\eta}}\left(2\cosh^2{\eta}-1\right)
\end{cases}
\end{equation}
together with the constraint
\be\label{eq_comp_hat_rho}
\begin{split}
& -\frac{5}{2}\phi^3\cosh^3{\eta} + 10m\phi^2\sin\beta\cosh^2{\eta} + \left[14m^2\cosh^3{\eta} -2m^2\left(\cos^2{\beta} + \frac{21}{2}\right)\cosh{\eta}\right]\phi +\\
& 24m^3\sin\beta\left(- \cosh^2{\eta} +\cos^2\beta+1\right)=0,
\end{split}
\ee
obtained by inserting expression \eqref{rho_LRSI} into eq. \eqref{eq_cov_LRSI_perfect12}. The constraint \eqref{eq_comp_hat_rho} is not preserved along the solutions of the system \eqref{equazioni_finali_LRSI}. Accordingly, we must implement a constraint algorithm, by looking for the points of the submanifold \eqref{eq_comp_hat_rho} where the dynamics \eqref{equazioni_finali_LRSI} is tangent to the submanifold itself. This produces an additional submanifold of \eqref{eq_comp_hat_rho}, described by a Cartesian equation (here omitted for brevity), for the unknowns $\phi$, $\beta$ and $\eta$. Unfortunately, we are not able to solve explicitly such an equation for any of its variables. Therefore, the constraint algorithm does not stabilize after the first step, but we are unable to proceed further due to purely computational reasons. We can conclude that there certainly are no solutions of eqs. \eqref{equazioni_finali_LRSI} and \eqref{eq_comp_hat_rho} in which all variables $\phi$, $\beta$ and $\eta$
are free. The constraint algorithm does not stabilize after the first step, so at
most only one variable would remain free. However, we are not able to prove with certainty
whether the problem admits solutions or not. We note that a similar conclusion can be reached also in the case $\eta=0$, as we discussed in \cite{VDFC}.
\subsection{LRSII space-times}
In these space-time classes, both the vorticity scalar $\Omega$ and the twist $\xi$ are zero. From eq. \eqref{constraint_H} we have immediately
\be
H=0.
\ee
Moreover, the covariant equations \eqref{dot_Omega}, \eqref{dot_H}, \eqref{dot_xi}, \eqref{hat_Omega}, \eqref{hat_xi} and \eqref{hat_H} are identically verified. Following \cite{VDFC}, the idea is to treat $Q$ and $\Pi$ as independent variables and use equations \eqref{momentum} and \eqref{anpressure} to obtain evolution and propagation equations for $\beta$. Indeed, from equations \eqref{momentum} and \eqref{anpressure}, we get the relations
\begin{subequations}\label{dot_hat_beta_LRSII-np}
\begin{align}
&\label{dotbeta_np_LRSII}\dot{\beta}=\frac{2}{\rho}\left[-2Q\cosh{\eta}+\left(3\Pi+m\rho\cos{\beta}\right)\sinh{\eta}\right],\\
&\label{hatbeta_np_LRSII}\hat{\beta}=\frac{2}{\rho}\left[-3\Pi\cosh{\eta}+2Q\sinh{\eta}\right].
\end{align}
\end{subequations}
Clearly, the integrability condition \eqref{inteffe} applied to $\beta$ 
\bea\label{intbeta} 
\hat{\dot{\beta}} - \dot{\hat{\beta}}= + A\dot{\beta} - \Sigma\hat{\beta} + \frac{1}{3}\Theta\hat{\beta} 
\eea
 must be verified. To check this fact, let us first observe that, by replacing equations \eqref{dotbeta_np_LRSII} and \eqref{hatbeta_np_LRSII} into equations \eqref{energy} and \eqref{pressure}, the energy density and pressure can be written as
\begin{subequations}\label{energy_pressure_np_LRSII}
\begin{align}
&\label{energy_np_LRSII}\mu=\frac{1}{2}\left(3\Pi+m\rho\cos{\beta}\right),\\
&\label{pressure_np_LRSII}p=\frac{1}{2}\Pi.
\end{align}
\end{subequations}
After that, a direct calculation shows that the integrability condition \eqref{intbeta} is ensured by the covariant equations \eqref{dot_mu} and \eqref{dot_Q}. In fact, by making the equation \eqref{intbeta} explicit and inserting the expressions \eqref{energy_pressure_np_LRSII} into equations \eqref{dot_mu} and \eqref{dot_Q}, the identity $\frac{\rho}{4}\eqref{intbeta}= \eqref{dot_mu}\cosh\eta - \eqref{dot_Q}\sinh\eta $ is easily proved.

In conclusion, the system of remaining covariant equations takes the form
\begin{subequations}\label{eq_finali_np_LRSII}
\begin{align}
&\dot{\Theta} + \hat{A} + \frac{3 }{2}\Pi + \frac{1}{3}\Theta^2 + \frac{3 }{2}\Sigma^2 + A \phi + \frac{1}{4} m \rho \cos{\beta}- A^2=0  \\
&\frac{3}{2} \dot{\Pi} - \hat{Q}  + 2 A Q + 2 \Pi \Theta +\frac{1}{2} \hat{\eta} m \rho \cos{\beta} + 2m Q \cosh{\eta} \sin{\beta} -\frac{3}{2} \Pi \Sigma - Q \phi - 3 m \Pi \sin{\beta} \sinh{\eta} = 0 \label{eq_finali_np_LRSII_b}\\
&\dot{Q}  - \frac{3}{2} \hat{\Pi}+ 3 A \Pi + \frac{4}{3} Q \Theta + \frac{1}{2} A m \rho \cos{\beta}-  Q \Sigma- \frac{3}{2} \Pi \phi =0 \label{eq_finali_np_LRSII_c} \\
&\dot{\Sigma}- \frac{2}{3}\hat{A}+\frac{1}{3}A\phi-\frac{1}{2}\Sigma^2+ \frac{2}{3}A^2+\frac{2}{3}\Theta\Sigma+E+\frac{1}{2}\Pi =0\\
&\dot{\phi}+\left(\Sigma+\frac{2}{3}\Theta\right)\left(A+\frac{1}{2}\phi\right)+Q=0 \\
&\nonumber\dot{E} + E\left(\Theta + \frac{3}{2}\Sigma\right) + \frac{1}{2}\Theta\Pi + Q\left(\frac{2}{3}m\sin\beta\cosh\eta - \frac{1}{2}\phi\right) +\frac{1}{4}\Sigma \left(m\rho\cos\beta\ + 3\Pi\right)+ \\
& \ \ \ \ \ \ \ \ \ \ \ \ \ \ \ \ \ \ \ \ \ \ 
+\tfrac16m\rho\hat{\eta}\cos\beta-m\Pi\sin\beta\sinh\eta=0 \\
&\dot{\ln{\rho}}-\hat{\eta}+\Theta-2m\sinh{\eta}\sin{\beta}=0 \label{eq_finali_np_LRSII_g}\\
&\dot{\eta} - \hat{\ln{\rho}} -\phi +A+2m\cosh{\eta}\sin{\beta}=0 \label{eq_finali_np_LRSII_h}\\
&\dot{\beta} -\frac{2}{\rho}\left[-2Q\cosh{\eta}+\left(3\Pi+m\rho\cos{\beta}\right)\sinh{\eta}\right]=0\\
&\hat{\Sigma} + \frac{2}{3}\hat{\Theta}+\frac{3}{2}\Sigma\phi+Q=0 \\
&\hat{\phi}+\frac{1}{2}\phi^2+\left(\Sigma-\frac{1}{3}\Theta\right)\left(\Sigma+\frac{2}{3}\Theta\right)+\frac{1}{3}m\rho\cos\beta 
+ \frac{3}{2}\Pi-E=0 \\
& \nonumber \hat{E} + Q (\frac{1}{3}\Theta + \frac{1}{2} \Sigma) + \frac{3}{2} \phi (E - \frac{1}{2}\Pi) +\frac{1}{6} m \rho \cos{\beta} (A + \dot{\eta}- \phi+ 2 m \cosh{\eta} \sin{\beta})+ m\Pi \sin{\beta}  \cosh{\eta}-\\ 
& \ \ \ \ \ \ \ \ \ \ \ \ \ \ \ \ \ \ \ \ \ \ - \frac{2}{3}m Q\sin{\beta}  \sinh{\eta}  = 0 \\
&\hat{\beta}-\frac{2}{\rho}\left[-3\Pi\cosh{\eta}+2Q\sinh{\eta}\right]=0.
\end{align}
\end{subequations}
From the system \eqref{eq_finali_np_LRSII}, further equations can be obtained by applying the integrability conditions \eqref{inteffe}. For instance, from eqs. \eqref{eq_finali_np_LRSII_g} and \eqref{eq_finali_np_LRSII_h} we get the wave-like equation
\begin{eqnarray}\label{intlnrhocondition}
\nonumber \ddot\eta\!-\!\hat{\hat\eta}&=&\left(\Sigma\!-\!\tfrac13\Theta\!-\!2m\sin\beta\sinh\eta\right)\dot\eta-\left(A\!-\!2m\sin\beta\cosh\eta\right)\hat{\eta}\!-\!\dot{A}\!-\!\hat\Theta\!-\!Q\!-\!\tfrac32\Sigma\phi+\\
 && \!+\!\tfrac23m\sin\beta\left[\left(3\Sigma\!-\!\Theta\right)\cosh\eta\!-\!3A\sinh\eta\right]\!+\!\frac{4m}{\rho}\cos\beta\left[2Q\cosh{2\eta}\!-\!\left(3\Pi\!+\!\tfrac12m\rho\cos\beta\right)\sinh{2\eta}\right],
\end{eqnarray}
while eqs. \eqref{eq_finali_np_LRSII_b} and \eqref{eq_finali_np_LRSII_c} yield
\begin{equation}\label{intPicondition}
\begin{aligned}
\ddot{Q} - \hat{\hat{Q}} &= 
(-5 A\! + \!2\phi \!-\! 2 m \sin\beta \cosh\eta)\hat{Q} 
\!+\! \left[3 (\Sigma \!-\!\Theta ) \!+ \!2 m \sin\beta \sinh\eta\right] \dot{Q} + \\
&\quad\!+\!(4 A^2 \!-\! 4 A\phi \!-\! 3 \Pi\! -\! \frac{14}{9} \Theta^2 \!+\! \frac{10}{3} \Theta \Sigma \!-\! \frac{1}{2} \Sigma^2 \!+\! \frac{1}{2} \Phi^2) Q \!-\! \frac{1}{6} m \rho (3 \dot{A} \!-\! 3 A \hat{\eta} \!+\! 3 \dot{\eta} \hat{\eta} \!+\! 3 \hat{\hat{\eta}} \!+\! 2 A \Theta \!-\! 6 A \Sigma) \cos\beta \\
&\quad \!+\! \frac{1}{2} m \sin\beta \left[\big( 3 \hat{\eta}\! +\! 2 A Q \!-\! 2 Q \Phi \!-\! \hat{\eta} m \rho \cos\beta \big) \cosh\eta \!+ \big( 3 A \Pi \!+\! 2 \hat{\eta} Q \!+\! \tfrac83 Q \Theta \!-\! 2Q \Sigma \!+\!  2\dot{Q} \sinh\eta \big) \sinh\eta\right] \\
&\quad\! +\!  \frac{m}{\rho} \left(4 Q \cos\beta \cosh\eta \!+\! \hat{\eta} \rho \sin\beta \!+\! 6 \Pi \cos\beta \sinh\eta\right) (2 Q \sinh\eta\!-\!3 \Pi \cosh\eta) \! +\! A m^2\rho \sin\beta  \cos\beta \sinh\eta .
\end{aligned}
\end{equation}
The system \eqref{eq_finali_np_LRSII} is composed of thirteen equations involving ten unknowns. Equations \eqref{eq_finali_np_LRSII} are strongly coupled and, above all, some of them contain both {\it dot} and {\it hat} derivatives simultaneously. This makes the known solution procedure for perfect fluids in LRSII space-times \cite{Ellisperf} inapplicable (also in the case $\eta=0$, as we erroneously wrote in \cite{VDFC}). The search for solution methods for the system \eqref{eq_finali_np_LRSII} deserves specific attention, and future research will be devoted to this topic. It is likely that solutions should be sought by assuming suitable simplifying hypotheses on some of the unknown functions. In this regard, an example is given below in the discussion of the perfect spinorial fluid case.

\subsubsection*{\bf Perfect fluid case}

In the case the spinorial fluid is perfect ($Q=\Pi=0$), from equations \eqref{dotbeta_np_LRSII} and  \eqref{hatbeta_np_LRSII} we have
\begin{subequations}\label{dothatbetaLRSII}
\begin{align}
&\dot{\beta}=2m\cos{\beta}\sinh{\eta},\\
&\hat{\beta}=0.
\end{align}
\end{subequations}
Analogously, equations \eqref{energy_np_LRSII} and \eqref{pressure_np_LRSII} imply
\begin{subequations}
\begin{align}
&\mu=\frac{1}{2}\rho m\cos{\beta}\label{energyLRSII}\\
&p=0\label{pressureLRSII},
\end{align}
\end{subequations}
which shows that in this case, the perfect spinorial fluid is necessarily a dust. Furthermore, by requiring that $\mu\not =0$, the vanishing of $\Omega$, $\xi$ and $p$ implies (see eq. \eqref{dot_Q}) 
\begin{eqnarray}\label{H_A_LRSII}
 A=0 .
\end{eqnarray}
The equations \eqref{dot_Omega}, \eqref{dot_H}, \eqref{dot_xi}, \eqref{hat_Omega}, \eqref{hat_xi} and \eqref{hat_H} are identically satisfied, whereas from equations \eqref{dot_mu}, \eqref{dotrho} and \eqref{energyLRSII} we get the condition
\be
\hat\eta =0.
\ee
Summing it all up, the covariant equations for the remaining undetermined unknowns $(\Theta,\Sigma,\phi,E,\rho,\beta,\eta)$ are expressed as
\begin{subequations}\label{fieleq}
\begin{align}
\label{dot_Theta_LRSII_perfect}
&\dot{\Theta}=-\dfrac{1}{3}\Theta^{2}-\dfrac{3}{2}\Sigma^{2}-\dfrac{1}{2}\mu \\
\label{dot_Sigma_LRSII_perfect}
&\dot{\Sigma}=\dfrac{1}{2}\Sigma^{2}-\dfrac{2}{3}\Theta\Sigma-E \\
\label{dot_phi_LRSII_perfect}
&\dot{\phi}=-\dfrac{1}{2}\phi\left(\Sigma+\frac{2}{3}\Theta\right) \\
\label{dot_E_LRSII_perfect}
&\dot{E}=-\Theta E - \dfrac{3}{2}E\Sigma - \dfrac{1}{2}\mu\Sigma \\
\label{hat_Sigma_LRSII_perfect}
&\hat{\Sigma}= -\dfrac{2}{3}\hat{\Theta}-\dfrac{3}{2}\Sigma\phi \\
\label{hat_phi_LRSII_perfect}
&\hat{\phi}=-\dfrac{1}{2}\phi^{2}-\left(\Sigma-\dfrac{1}{3}\Theta\right)\left(\Sigma+\dfrac{2}{3}\Theta\right)-\dfrac{2}{3}\mu+E \\
\label{hat_E_LRSII_perfect}
&\hat{E}= - \dfrac{1}{3}\hat{\mu}-\dfrac{3}{2}E\phi \\
\label{dot_rho_LRSII_perfect}
&\dot{\ln\rho}=-\Theta+2m\sinh{\eta}\sin{\beta} \\
\label{dot_beta_LRSII_perfect}
&\dot\beta=2m\sinh\eta\cos\beta \\
\label{dot_eta_LRSII_perfect}
&\dot{\eta}=\hat{\ln\rho}+\phi-2m\cosh{\eta}\sin{\beta} \\
\label{hat_beta_LRSII_perfect}
&\hat\beta=0 \\
\label{hat_eta_LRSII_perfect}
&\hat\eta=0,
\end{align}
\end{subequations}
where $\mu$ is given by eq. \eqref{energyLRSII}. In particular and as expected, eqs. \eqref{dot_rho_LRSII_perfect} and \eqref{dot_beta_LRSII_perfect} ensure that the mass-energy density $\mu$ undergoes the standard conservation law of a dust
\begin{equation}\label{dust_law}
\dot\mu=-\Theta\mu.
\end{equation}
The consistency and integrability of equations \eqref{dot_Theta_LRSII_perfect}-\eqref{hat_E_LRSII_perfect}, along with eq. \eqref{dust_law}, have been discussed in \cite{VDFC} and are ensured by the conditions \eqref{rif1} and \eqref{rif2}. As already proven, eqs. \eqref{dot_beta_LRSII_perfect} and \eqref{hat_beta_LRSII_perfect} satisfy the commutation relation \eqref{inteffe}, whereas the same relation \eqref{inteffe}, applied to equations \eqref{dot_eta_LRSII_perfect} and \eqref{hat_eta_LRSII_perfect}, yields the additional equation
\be\label{additional_equation_rho_LRSII_perfect}
\hat{\hat{\ln\rho}} + \hat\phi=0.
\ee 

Assuming $\mu$ as a free variable, eqs. \eqref{dot_Theta_LRSII_perfect}-\eqref{hat_E_LRSII_perfect} and \eqref{dust_law} result to be decoupled from the remaining ones \eqref{dot_rho_LRSII_perfect}-\eqref{hat_eta_LRSII_perfect} and \eqref{additional_equation_rho_LRSII_perfect}. The first set of equations (\eqref{dot_Theta_LRSII_perfect}-\eqref{hat_E_LRSII_perfect} and \eqref{dust_law}) describes an LRSII space-time filled by a dust. Exact solutions of such equations have been widely discussed in the literature (see, for example, \cite{Ellisperf,Ellis_67} and references therein). A general algorithm for solving eqs. \eqref{dot_Theta_LRSII_perfect}-\eqref{hat_E_LRSII_perfect} and \eqref{dust_law} consists in freely choosing initial data for the unknowns $\Theta$ and $\mu$ on a space-like hypersurface $\sigma$, and then using the propagation equations \eqref{hat_Sigma_LRSII_perfect}, \eqref{hat_phi_LRSII_perfect} and \eqref{hat_E_LRSII_perfect} to determine the spatial distribution of $\Sigma$, $\phi$ and $E$ on $\sigma$; after that, eqs. \eqref{dot_Theta_LRSII_perfect}-\eqref{dot_E_LRSII_perfect} provide us with the evolution laws of the unknowns $\Theta$, $\Sigma$, $\phi$ and $E$ along the time-like congruence. Making use of the solutions for $\Theta$ and $\phi$ so found, it would then be a matter of solving the second set of equations \eqref{dot_rho_LRSII_perfect}-\eqref{hat_eta_LRSII_perfect} and \eqref{additional_equation_rho_LRSII_perfect}. Unfortunately, the presence of the additional equation \eqref{additional_equation_rho_LRSII_perfect} prevents this general procedure from being applied, since \eqref{additional_equation_rho_LRSII_perfect} is not preserved along the time-like congruence. Anyway, particular solutions can be investigated by assuming some simplifying assumptions, for example: 1) $E=0$, 2) $\Sigma=0$, 3) $\phi=0$.  
\\
\\
1) $E=0$.

Still assuming $\mu\not=0$, from the equations (\ref{dot_E_LRSII_perfect}), (\ref{hat_E_LRSII_perfect}) and (\ref{hat_beta_LRSII_perfect}) we get the conditions
\begin{equation}\label{condizioni_Sigma_rho_LRSII_perfect_E0}
\Sigma=0 \quad {\rm and} \quad \hat{\rho}=0,
\end{equation}
so that the system \eqref{fieleq} reduces to
\begin{subequations}\label{fieleqE0}
\begin{align}
\label{dot_theta_LRSII_E0}
&\dot{\Theta}=-\frac{1}{3}\Theta^{2}-\frac{1}{2}\mu \\
\label{dot_phi_LRSII_E0}
&\dot{\phi}=-\frac{1}{3}\phi\Theta \\
\label{dot_lnrho_LRSII_E0}
&\dot{\ln{\rho}}=-\Theta+2m\sinh{\eta}\sin{\beta} \\
\label{doteta_LRSII_E0}
&\dot{\eta}=\phi-2m\cosh{\eta}\sin{\beta} \\
\label{dotbeta_LRSII_E0}
&\dot{\beta}=2m\cos{\beta}\sinh{\eta} \\
\label{hat_theta_LRSII_E0}
&\hat{\Theta}=0 \\
\label{hat_phi_LRSII_E0}
&\hat{\phi}=-\frac{1}{2}\phi^{2}+\frac{2}{9}\Theta^2-\frac{2}{3}\mu \\
\label{hat_rho_LRSII_E0}
&\hat\rho=0\\
\label{hat_beta_LTSII_E0}
&\hat\beta=0\\
\label{hat_eta_LRSII_E0}
&\hat\eta=0.
\end{align}
\end{subequations}
Moreover, due to eq. \eqref{hat_rho_LRSII_E0}, eq. \eqref{additional_equation_rho_LRSII_perfect} simplifies to
\be\label{hat_phi_LRSII_E0_bis}
\hat\phi=0.
\ee
The space-time and the Dirac field resulting from equations \eqref{fieleqE0} are then homogeneous. They are characterized by the final set of differential equations
\begin{equation}\label{final_set_LRSII_perfect_E0}
\begin{cases}
&\dot{\Theta}=-\frac{1}{3}\Theta^{2}-\frac{1}{4}m\rho\cos\beta \\
&\dot{\phi}=-\frac{1}{3}\phi\Theta \\
&\dot{\rho}=\rho\left(-\Theta+2m\sinh{\eta}\sin{\beta}\right) \\
&\dot{\eta}=\phi-2m\cosh{\eta}\sin{\beta} \\
&\dot{\beta}=2m\cos{\beta}\sinh{\eta}
\end{cases}
\end{equation}
together with the equation
\be\label{idconstraint_LRSII_E0}
\frac{1}{2}\phi^{2}-\frac{2}{9}\Theta^2+\frac{1}{3}m\rho\cos\beta =0,
\ee
coming from eq. \eqref{hat_phi_LRSII_E0} and representing a constraint on the initial data. Indeed, making use of equations \eqref{energyLRSII}, \eqref{dot_theta_LRSII_E0}, \eqref{dot_phi_LRSII_E0}, \eqref{dot_lnrho_LRSII_E0} and \eqref{dotbeta_LRSII_E0}, a direct check shows that the constraint \eqref{idconstraint_LRSII_E0} is preserved by the dynamics \eqref{final_set_LRSII_perfect_E0}.

We note that a particular case of \eqref{final_set_LRSII_perfect_E0} is obtained by setting $\phi=0$. In such a circumstance, the field equations \eqref{final_set_LRSII_perfect_E0} and \eqref{idconstraint_LRSII_E0} become
\begin{equation}\label{fieleqEphi0}
\begin{cases}
&\dot{\Theta}=-\frac{1}{2}\Theta^{2} \\
&\frac{1}{3}\Theta^2=\frac{1}{2}m\rho\cos\beta \\
&\dot{\rho}=\rho\left(-\Theta+2m\sinh{\eta}\sin{\beta}\right)\\
&\dot{\eta}=-2m\cosh{\eta}\sin{\beta}\\
&\dot{\beta}=2m\cos{\beta}\sinh{\eta}.
\end{cases}
\end{equation}
The system (\ref{fieleqEphi0}) describes a spinorial dust in a homogeneous, isotropic, and conformally flat space-time. For $\eta=0$ and $\beta=0$, the system (\ref{fieleqEphi0}) admits a specific solution expressed in comoving coordinates as a spatially flat FLRW space-time \cite{VDFC}. A more general integration of eqs. (\ref{fieleqEphi0}) is illustrated in Section \ref{sezione_8}.
\\
\\
2) $\Sigma=0$.

The assumption $\Sigma=0$, via the evolution equation (\ref{dot_Sigma_LRSII_perfect}), entails $E=0$, thus falling back into case 1).
\\
\\
3) $\phi=0$.

The evolution equation (\ref{dot_phi_LRSII_perfect}) is automatically satisfied, meanwhile the propagation equation (\ref{hat_phi_LRSII_perfect}) yields the constraint
\be\label{constraint_phi0}
E=\left(\Sigma-\frac{1}{3}\Theta\right)\left(\Sigma+\frac{2}{3}\Theta\right)+\frac{2}{3}\mu.
\ee
The remaining equations are
\begin{subequations}\label{fieleqphi0}
\be\label{dot_theta_LRSII_phi0}
\dot{\Theta}=-\frac{1}{3}\Theta^{2}-\frac{3}{2}\Sigma^{2}-\frac{1}{2}\mu
\ee
\be\label{dot_Sigma_LRSII_phi0}
\dot{\Sigma}=\frac{1}{2}\Sigma^{2}-\frac{2}{3}\Theta\Sigma-\left(\Sigma-\frac{1}{3}\Theta\right)\left(\Sigma+\frac{2}{3}\Theta\right)-\frac{2}{3}\mu
\ee
\be\label{dot_E_LRSII_phi0}
\dot{E}=- \Theta E - \frac{3}{2}E\Sigma - \frac{1}{2}\mu\Sigma 
\ee
\be\label{dot_lnrho_LRSII_phi0}
\dot{\ln{\rho}}=-\Theta+2m\sinh{\eta}\sin{\beta}
\ee
\be\label{hat_lnrho_LRSII_phi0}
\dot{\eta} = \hat{\ln{\rho}} - 2m\cosh{\eta}\sin{\beta}
\ee
\be\label{dot_beta_LRSII_phi0}
\dot\beta = 2m\sinh\eta\cos\beta
\ee
\be\label{hat_Sigma_LRSII_phi0}
\hat{\Sigma}+\frac{2}{3}\hat{\Theta}=0
\ee
\be\label{hat_E_LRSII_phi0}
\hat{E}+\frac{1}{3}\hat{\mu}=0
\ee
\be
\hat{\hat{\ln{\rho}}} =0
\ee
\be\label{hat_beta_LRSII_phi0}
\hat\beta=0
\ee
\be\label{hat_eta_LRSII_phi0}
\hat\eta=0.
\ee
\end{subequations}
Expression (\ref{constraint_phi0}) satisfies the evolution equation (\ref{dot_E_LRSII_phi0}), instead the consistency with (\ref{hat_E_LRSII_phi0}) gives rise to the additional equation
\be\label{add_eq_LRSII_phi0}
\hat{\Theta}\left(\Sigma+\frac{2}{3}\Theta\right)-\frac{1}{2}m\hat{\rho}\cos{\beta}=0.
\ee
Once again, the problem can be simplified by requiring homogeneity ($\hat{f}=0$ for every covariantly defined scalar function $f$). In fact, under such an assumption the field equations \eqref{fieleqphi0} and \eqref{add_eq_LRSII_phi0} reduce to 
\begin{equation}\label{fieleqphi0homo}
\begin{cases}
&\dot{\Theta}=-\frac{1}{3}\Theta^{2}-\frac{3}{2}\Sigma^{2}-\frac{1}{4}m\rho\cos\beta \\
&\dot{\Sigma}=\frac{1}{2}\Sigma^{2}-\frac{2}{3}\Theta\Sigma-\left(\Sigma-\frac{1}{3}\Theta\right)\left(\Sigma+\frac{2}{3}\Theta\right)-\frac{1}{3}m\rho\cos\beta \\
&\dot{\rho}=\rho\left(-\Theta+2m\sinh{\eta}\sin{\beta}\right) \\
&\dot{\eta}=-2m\cosh{\eta}\sin{\beta} \\
&\dot{\beta}=2m\cos{\beta}\sinh{\eta} .
\end{cases}
\end{equation}
For $\eta=0$ and $\beta=0$, the solution of the system \eqref{fieleqphi0homo} describes a Bianchi-I space-time filled by a spinorial dust. Indeed, in such a circumstance, a direct check shows that the first two equations of \eqref{fieleqphi0homo} together with eq. \eqref{constraint_phi0} result to be suitable linear combinations of Einstein's equations obtained in co-moving coordinates \cite{VDFC}.
\subsection{LRSIII space-times}
We now focus on twisting and non-rotating geometries i.e. $\xi\neq0$, $\Omega=0$ and $Q=0$.
Eqs. \eqref{dot_Omega}, \eqref{rif1} and \eqref{inteffe} imply $\hat{f}=0$ (for every covariantly defined scalar function $f$), $\phi=0$ and $A=0$. Due to the vanishing of $Q$, condition \eqref{perfectfluidconstr1} yields the evolution equation for $\beta$
\be\label{dotbeta_LRSIII_np}
\dot{\beta}=2m\cos{\beta}\sinh{\eta}-\xi.
\ee
In view of eq. \eqref{dotbeta_LRSIII_np}, the thermodynamic quantities can be expressed as
\begin{subequations}\label{thermo_LRSIII_np}
\begin{align}
&\label{energy_LRSIII_np} \mu=\frac{1}{4}\rho\left(2m\cos{\beta}+\xi\sinh{\eta}\right),\\
&\label{pression_LRSIII_np}p=\frac{1}{12}\xi\rho\sinh{\eta},\\
&Q=0,\\
&\label{anisotropic_pressure_LRSIII_np}\Pi=-\frac{1}{3}\xi\rho\sinh{\eta}.
\end{align}
\end{subequations}
The expression for the magnetic part $H$ of the Weyl tensor is given by eq. \eqref{constraint_H}, that is
\be\label{H_LRSIII_np}
H=-3\xi\Sigma,
\ee
whereas, from eq. \eqref{hat_phi}, we deduce the following expression for the electric part $E$ of the Weyl tensor 
\be\label{E_LRSIII_np}
E=\frac{1}{2}\Pi+\frac{2}{3}\mu+\left(\Sigma-\frac{1}{3}\Theta\right)\left(\Sigma+\frac{2}{3}\Theta\right)-2\xi^2.
\ee
All remaining propagation equations are identically verified. Furthermore, a direct calculation shows that the expression \eqref{H_LRSIII_np} and \eqref{E_LRSIII_np}, together with equations \eqref{dotbeta_LRSIII_np} and  \eqref{thermo_LRSIII_np} make the covariant equations  \eqref{dot_H}, \eqref{dot_mu}, \eqref{dot_E} automatically satisfied. All the remaining covariant equations, which are not automatically satisfied, can be collected in the following system of differential equations
\begin{eqnarray}\label{eq_finali_LRSIII_np}
\begin{cases}
&\dot{\xi}=-\dfrac{1}{3}\xi\left(\Theta+6\Sigma\right)\\[6pt]
&\dot{\Theta}=-\dfrac{1}{3}\Theta^2-\dfrac{3}{2}\Sigma^2-\dfrac{1}{4}m\rho\cos{\beta}-\frac{1}{4}\rho\xi\sinh{\eta}\\[6pt]
&\dot{\Sigma}=-\dfrac{1}{2}\Sigma^2-\Theta\Sigma+\dfrac{2}{9}\Theta^2+2\xi^2-\dfrac{1}{3}m\rho\cos{\beta}+\dfrac{1}{6}\rho\xi\sinh{\eta}\\[6pt]
&\dot{\rho}=\rho\left(2m\sinh{\eta}\sin{\beta}-\Theta\right)\\[6pt]
&\dot{\eta}=-2m\cosh{\eta}\sin{\beta}\\[6pt]
&\dot{\beta}=2m\sinh{\eta}\cos{\beta}-\xi.
\end{cases}
\end{eqnarray}
The system \eqref{eq_finali_LRSIII_np} contains six differential equations in normal form for the six unknowns $\{\xi,\Theta,\Sigma,\rho,\eta,\beta\}$. Thus, assigned initial data on a space-like hypersurface $\sigma$ orthogonal to $v^i$, the corresponding Cauchy problem admits a unique solution, at least locally. A numerical solution of eqs. \eqref{eq_finali_LRSIII_np} is shown in Section \ref{sezione_8}.

\subsubsection*{\bf Perfect fluid case}

In this case, eq. \eqref{dothatbeta2} implies necessarily $\eta=0$, so that the identities $v^i=u^i$ and $e^i=s^i$ automatically follow. Under these conditions, the problem has already been studied in \cite{VDFC}, where the absence of solutions was proved.

\section{Some examples of solutions}\label{sezione_8}
In this section, we derive and analyze both exact and numerical solutions of the differential systems introduced in the previous section. In particular, numerical methods are employed to investigate more complex geometries. 
\subsection{An exact solution}
We consider the system \eqref{fieleqEphi0} which describes a spinorial dust filling a LRSII isotropic, homogeneous, and conformally flat ($E=H=0$) space-time.
\begin{eqnarray}\label{fieleqEphi0-}
\begin{cases}
\dot{\Theta} = -\dfrac{1}{2}\Theta^2 \\[6pt]
\dfrac{1}{3}\Theta^2 = \dfrac{1}{2} m \rho \cos\beta \\[6pt]
\dot{\rho} = \rho(-\Theta + 2m\sinh\eta\sin\beta) \\[6pt]
\dot{\eta} = -2m\cosh\eta\sin\beta \\[6pt]
\dot{\beta} = 2m\cos\beta\sinh\eta.
\end{cases}
\end{eqnarray}
To solve the system \eqref{fieleqEphi0-}, we preliminarily observe that the first equation is decoupled from the others and can be integrated as 
\begin{equation}\label{thetasol}
\Theta = \frac{2}{t + C},
\end{equation}
where $C$ is an integration constant and $t$ denotes an affine parameter along the time-like congruence. The behavior of the fermion field, however, is non-trivial. To understand it, let us focus on the sub-system given by the last two equations of \eqref{fieleqEphi0-}, namely
\begin{eqnarray}\label{eq:system}
\begin{cases}
\dot{\eta} = -2m\cosh\eta\,\sin\beta\\[4pt]
\dot{\beta} = 2m\cos\beta\,\sinh\eta.
\end{cases}
\end{eqnarray}
Directly from eqs. \eqref{eq:system} we have the relation
\begin{equation}
\frac{d\beta}{d\eta}=-\tanh\eta\cot\beta,
\end{equation}
which implies the identity
\begin{equation}\label{constraint_beta_eta}
\cos\beta=k\cosh\eta,
\end{equation}
where
\begin{equation}\label{eq:invariant}
k:=\frac{\cos\beta_0}{\cosh\eta_0}, \qquad 0\leq |k| \leq 1
\end{equation}
is a constant determined by the initial values $\eta_0:=\eta(t_0)$ and $\beta_0:=\beta(t_0)$. As for the constant $k$, we may exclude the cases $|k|=0$ and $|k|$=1. Indeed, $|k|=0$ implies \(\cos\beta= 0\)
(\(\beta=\tfrac{\pi}{2}+n\pi\) constant): in this case, the mass-energy density vanishes identically, $\mu=0$. Instead, condition $|k|=1$ entails necessarily \(\eta=0\) and \(\beta= n\pi\): such a case has already been studied in detail in \cite{VDFC}. Accordingly, we assume hereafter that the initial data are chosen so that $0<|k|<1$.

In view of eq. \eqref{constraint_beta_eta}, we have also
\begin{equation}\label{sinbeta}
\sin\beta=s\sqrt{1-\cos^2\beta}=s\sqrt{1-k^2\cosh^2\eta},
\end{equation}
where \(s:=\operatorname{sign}(\sin\beta)\) is again determined by the initial data. Inserting eq. \eqref{sinbeta} into the first of eqs. \eqref{eq:system}, we get the final differential equation for the unknown $\eta$
\begin{equation}\label{eta_eq}
\dot{\eta}=-2m\cosh\eta\; s\sqrt{1-k^2\cosh^2\eta}.
\end{equation}
Equation \eqref{eta_eq} is valid as long as \(\sin\beta\neq0\) ($\beta \not= n\pi$), i.e. as long as the sign \(s\) is constant.  In these intervals of time, the above equation admits an implicit solution of the form
\begin{equation}\label{eq:quadrature}
t-t_0=-\frac{1}{2m\,s}\int_{\eta_0}^{\eta}
\frac{d\sigma}{\cosh\sigma\sqrt{1-k^2\cosh^2\sigma}}.
\end{equation}
An alternative expression for the solution \eqref{eq:quadrature} is obtained
by the substitution \(u=\cosh\sigma\):
\begin{equation}\label{quadrature_sub}
t-t_0=-\frac{1}{2m\,s}\int_{u_0}^{u}
\frac{du}{u\sqrt{u^2-1}\sqrt{1-k^2 u^2}}.
\end{equation}
The integral in the eq. \eqref{quadrature_sub} can be solved in terms of elliptic functions, giving rise to the explicit solution 
\begin{equation}\label{sol_eta_elliptic}
\eta=\operatorname{arccosh}\!\left(\frac{1}{k}\,\mathrm{sn}(f;k)\right),
\end{equation}
where \(\mathrm{sn}\) denotes the (Jacobi) elliptic sine and \(f\) is an affine function of \(t\). Once \(\eta\) is obtained, \(\beta\) is provided by eq. \eqref{constraint_beta_eta}
\begin{equation}
\beta=\arccos\bigl(k\cosh\eta\bigr).
\end{equation}
The function \(\eta\) is bounded and periodic, so $\beta$ is bounded and periodic too. In particular, due to Eq. \eqref{constraint_beta_eta}, we have  
\begin{equation}
|\eta|\le\operatorname{arccosh}\!\bigl(1/|k|\bigr).
\end{equation}
When $\sin\beta=0$, the first of \eqref{eq:system} implies $\dot\eta=0$ and eq. \eqref{constraint_beta_eta} implies that $\eta$ reaches the (extremum) value $\eta_*$ such that 
\begin{equation}
1-k^2\cosh^2\eta_*=0.
\end{equation} 
This result allows us to construct the full solution for $\eta$ by combining the integral \eqref{eq:quadrature} when  $\sin\beta\neq0$ and the value  $\eta = \eta_*$ when  $\sin\beta=0$.

Making use of the above results, we are able to implement and solve the equation for $\rho$
\begin{equation}
\dot{\rho} = -\rho(\Theta - 2m\sinh\eta\sin\beta).
\end{equation}
By separation of variables and taking eq. \eqref{sinbeta} into account, we get
\begin{equation}\label{rhosol}
\rho = \frac{B}{(t+C)^2\,k\cosh\eta},
\end{equation}
where $B$ is an integration constant whose value is singled out by requiring that the solution satisfies the constraint
\begin{equation}
\frac{1}{3}\Theta^2 = \frac{1}{2}m\rho\cos\beta,
\end{equation}
namely
\begin{equation}
B = \frac{8}{3m}.
\end{equation}
Summing it all up, the general solution of the system \eqref{fieleqEphi0-} is given by 
\begin{subequations}\label{final_solution_LRSII_perfect_Homogeneous_Isotropic}
\be
\Theta = \dfrac{2}{t+C} 
\ee
\be
\rho = \dfrac{8}{3mk(t+C)^2}\,\operatorname{sech}\eta 
\ee
\be
{\int_{\eta_0}^{\eta}\dfrac{du}{\cosh u\,\sqrt{1 - k^2\cosh^2 u}}} = -2m\,s\,(t - t_0) 
\ee
\be
\beta =\arccos\left(k\,\cosh\eta\right) ,
\ee
\end{subequations}
where the constants $k$ and $s$ are determined by the initial values $\eta_0$ and $\beta_0$.

\begin{figure}[H]
    \centering
    \begin{minipage}[b]{0.45\textwidth}
        \centering
        \includegraphics[height=4.5cm]{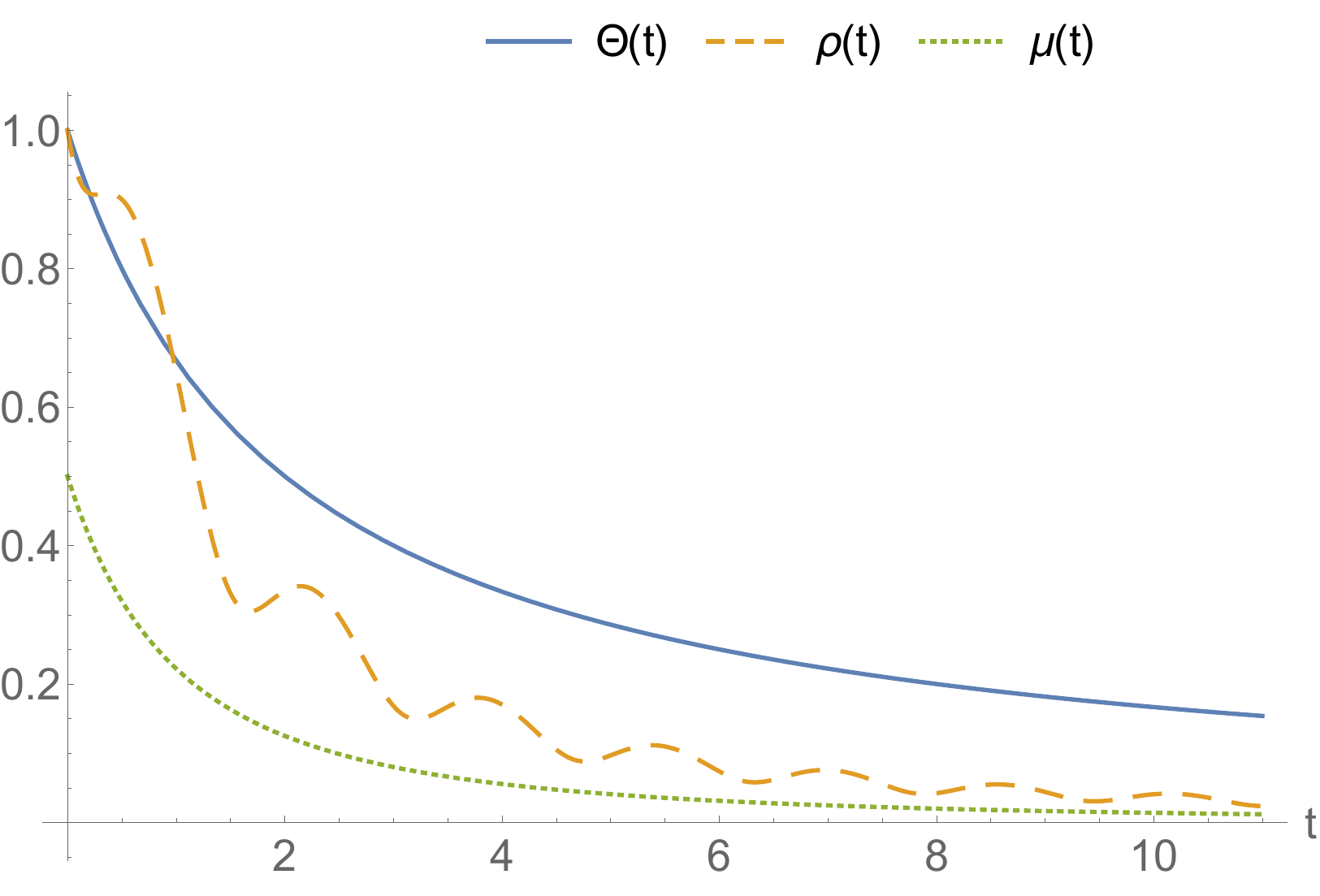}
    \end{minipage}
    \hspace{0.05\textwidth}
    \begin{minipage}[b]{0.45\textwidth}
        \centering
        \includegraphics[height=4.5cm]{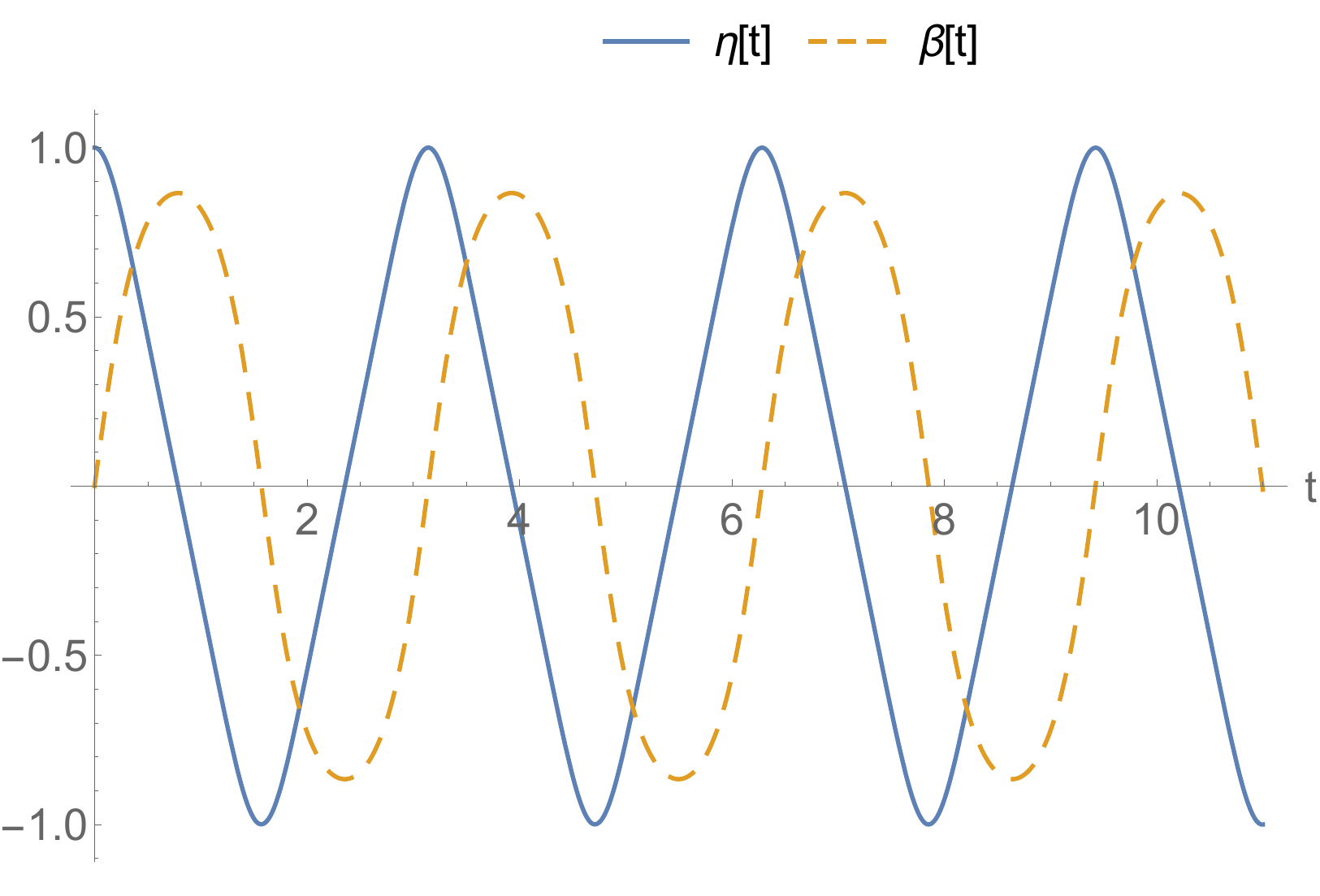}
    \end{minipage}
  \caption{\small Time evolution of the solution \eqref{final_solution_LRSII_perfect_Homogeneous_Isotropic} with  
initial data: $m=1$, $\Theta(0)=1$, $\rho(0)=1$, $\eta(0)=1$ and $\beta(0)=0$. 
On the left-hand side: corresponding behavior of the energy density $\mu$ (dotted line), the modulus $\rho$ (dashed line)
and the expansion scalar $\Theta$ (solid line). 
On the right-hand side: evolution of $\eta$ (solide line) and $\beta$ (dashed line).}
\label{fig:FLRW}
\end{figure}

The behavior of the exact solution \eqref{final_solution_LRSII_perfect_Homogeneous_Isotropic} is plotted in 
Fig.~\ref{fig:FLRW}. All quantities are expressed in dimensionless form by rescaling with appropriate characteristic scales. In natural units, the particle mass $m$ provides a natural unit for mass and energy, while the corresponding Compton length $\lambda_C = 1/m$ serves as a natural unit of length. Expressing variables in these units allows the equations to be written in dimensionless form. 
The figure shows that the functions $\rho$, $\eta$ and $\beta$ are indeed oscillating while  $\Theta$ and $\mu$ decreases in time. Note how the oscillation of $\beta$ and $\rho$ exactly compensate each other to produce the non-oscillating effective source. This mirrors the fact that only some aspects of the evolution of the spinor field influence the spacetime geometry. In this case, even if the spinor field exhibits oscillatory behavior, the net effect remains that of a pressureless fluid.

\subsection{Numerical solutions}
\subsubsection{Non-perfect spinorial fluid in LRSI space-times}

\begin{figure}[H]

    \centering
    \begin{minipage}[b]{0.45\textwidth}
        \centering
        \includegraphics[height=4.5cm]{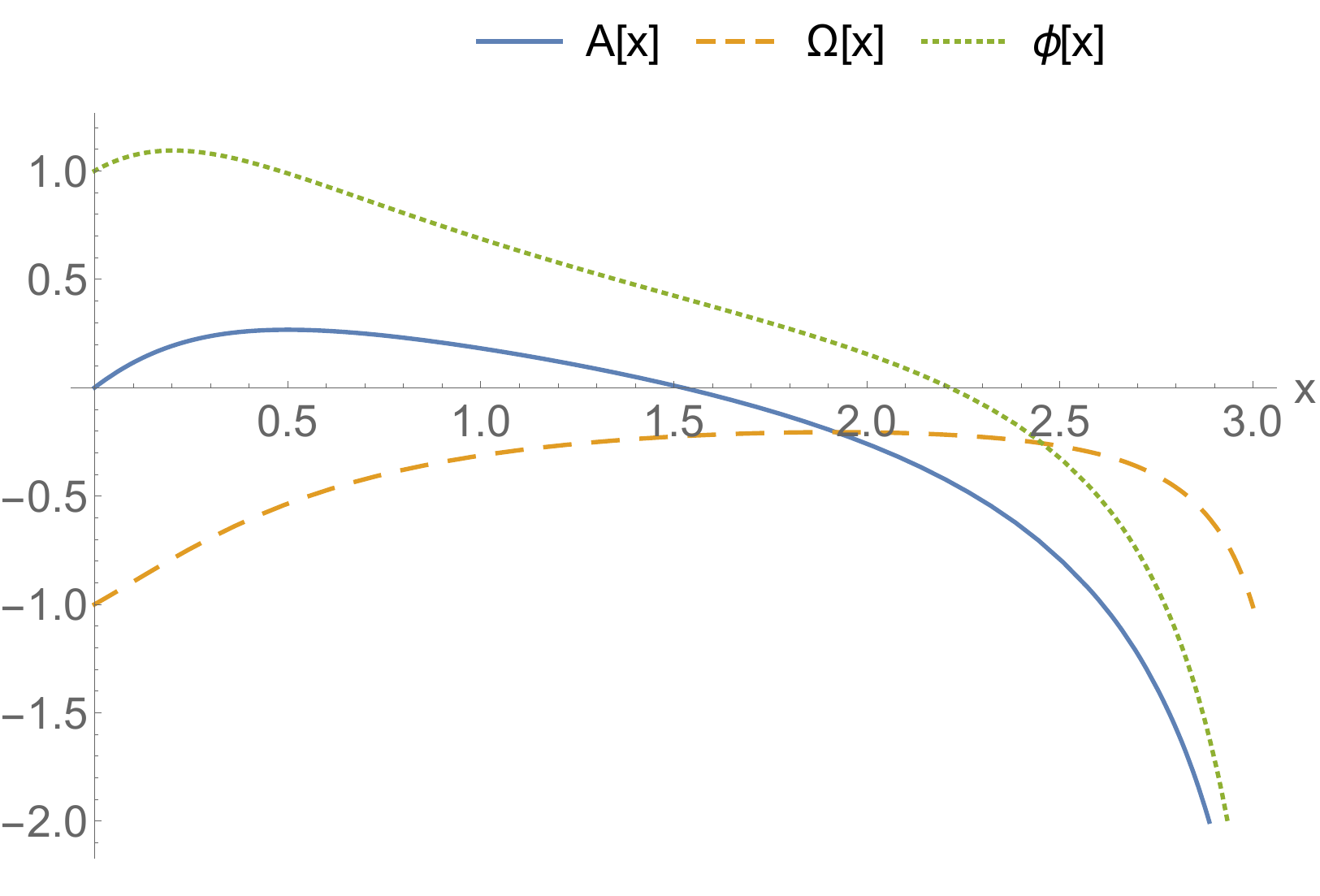}
    \end{minipage}
    \hspace{0.05\textwidth}
    \begin{minipage}[b]{0.45\textwidth}
        \centering
        \includegraphics[height=4.5cm]{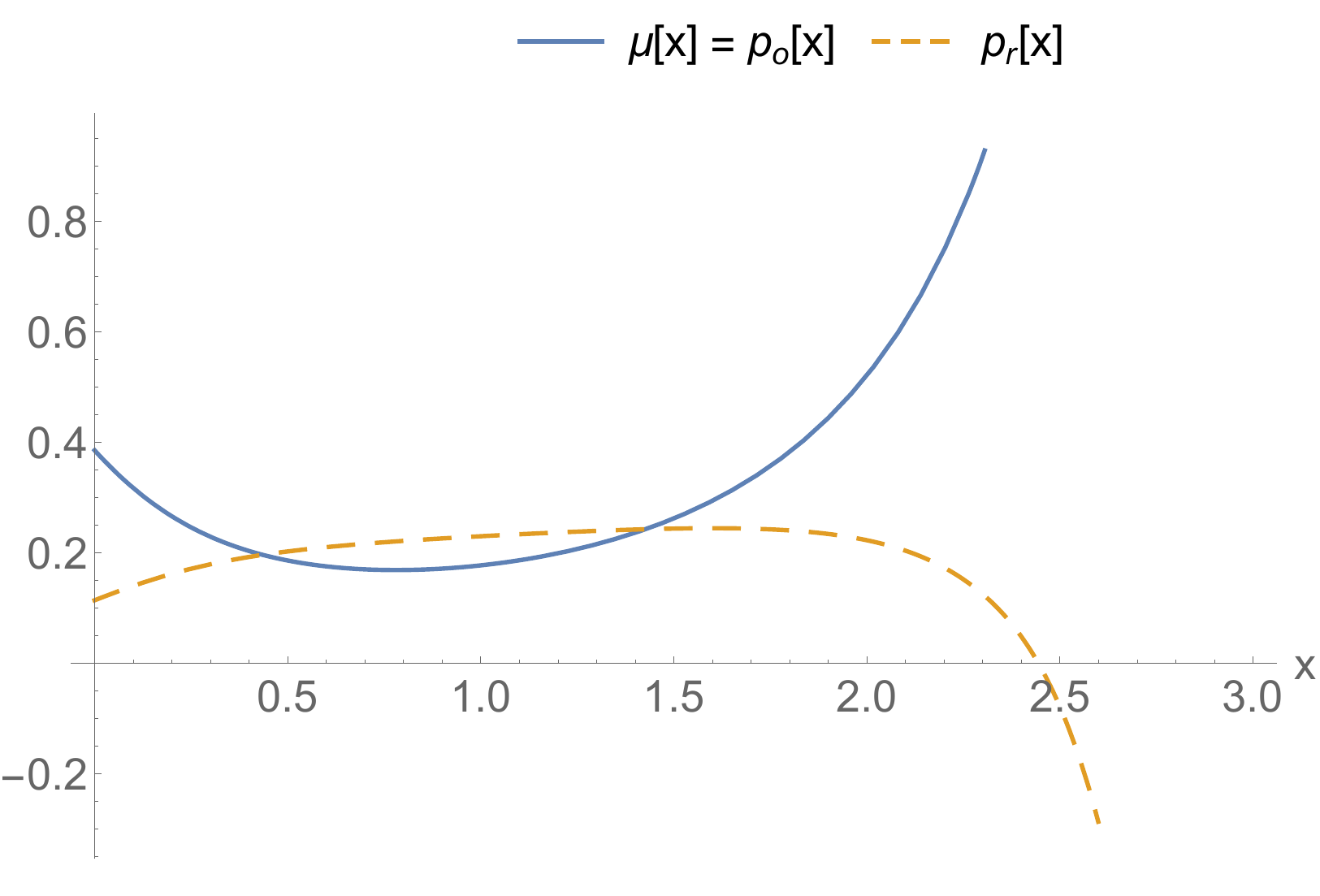}
    \end{minipage}
\caption{\small Evolution of the system \eqref{eq_finali_LRSI_non_perfect_} starting from the initial data $m = 1$, $A_0 = 0$, $\Omega_0 = -1$, $\phi_0 = \rho_0 = \eta_0 = 1$ and $\beta_0 = \pi$. Left-hand panel: behavior of the kinematical variables $A$ (solid line), $\phi$ (dotted line), and $\Omega$ (dashed line) in function of $x$. Right-hand panel: behavior of the energy density $\mu$ (solid line), radial pressure $p_r$ (dashed line), and orthogonal pressure $p_o$ (dotted line) as functions of $x$. The plots of $\mu$ and $p_o$ coincide.}
\label{Fig:LRSI1}
  \end{figure}

\begin{figure}[H]
\centering
    \begin{minipage}[b]{0.45\textwidth}
        \centering
        \includegraphics[height=5.0cm]{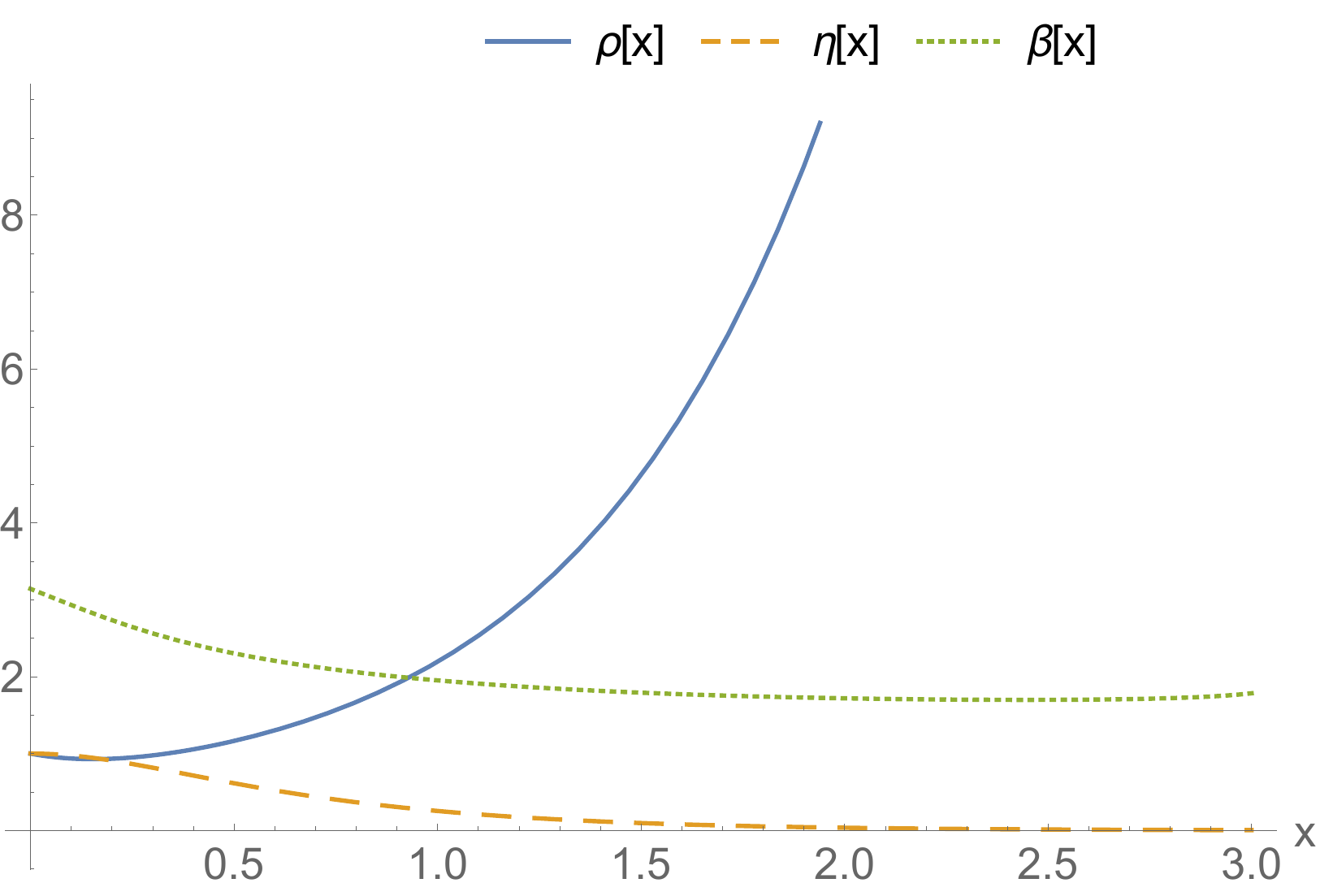}
  \end{minipage}
    \hspace{0.05\textwidth}
    \begin{minipage}[b]{0.45\textwidth}
        \centering
        \includegraphics[height=5.0cm]{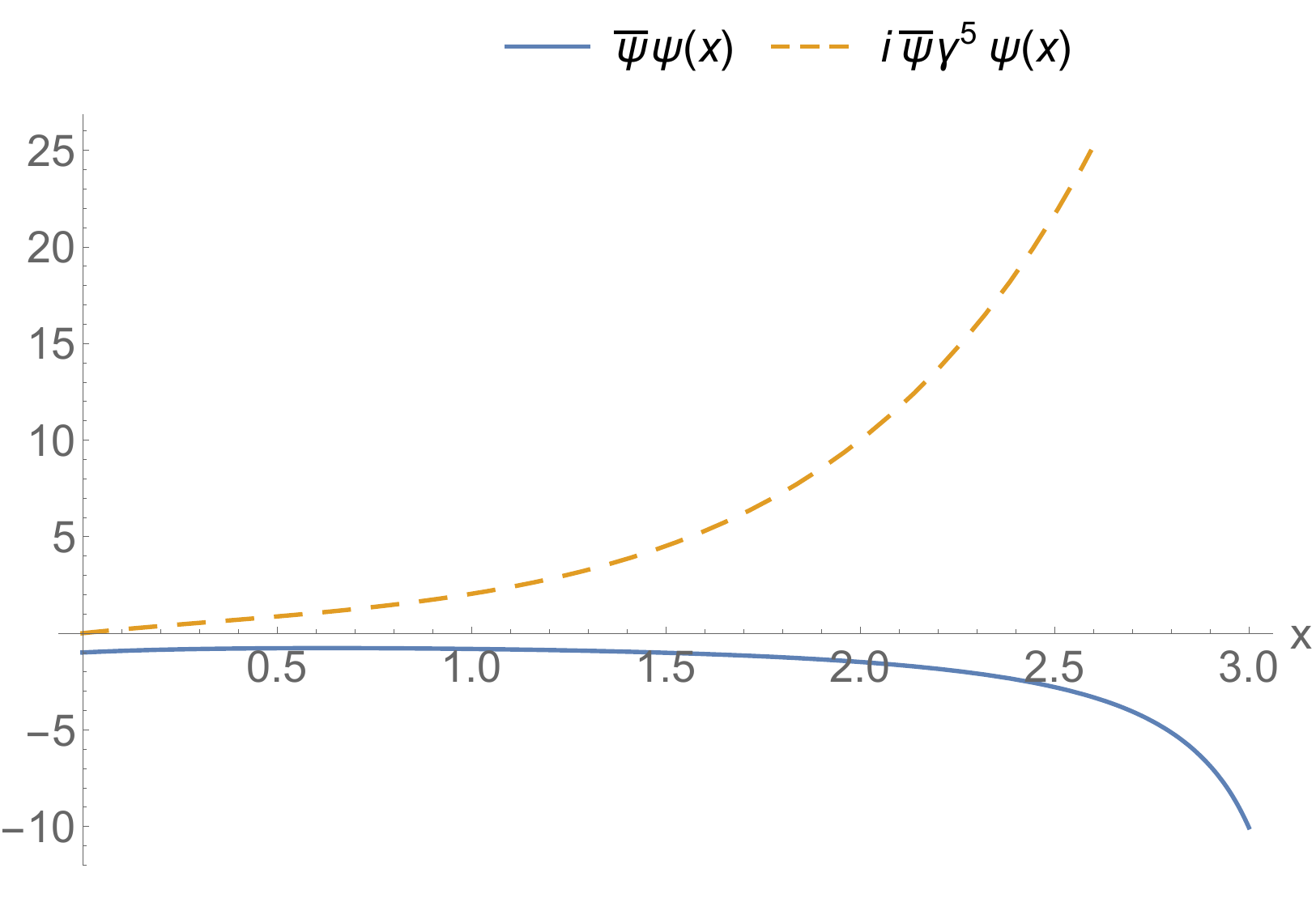}
    \end{minipage}
        \caption{\small Evolution of the system \eqref{eq_finali_LRSI_non_perfect_} with initial data $m = 1$, $A_0 = 0$, $\Omega_0 = -1$, $\phi_0 = \rho_0 = \eta_0 = 1$ and $\beta_0 = \pi$. Left-hand panel: dynamical behavior of the modulus $\rho$ (solid line), the pseudo-scalar function $\eta$ (dashed line) and the chiral angle $\beta$ (dotted line) in function of $x$.  Right-hand panel: behavior of the bilinear scalar $\bar{\psi}\psi$ (solid line) and the pseudo-scalar $i\bar{\psi}\gamma^5\psi$ (dashed line) as functions of $x$.}
\label{Fig:LRSI2}
\end{figure}

By using numerical techniques, we analyze the system \eqref{eq_finali_LRSI_non_perfect}
\bea\label{eq_finali_LRSI_non_perfect_}
\begin{cases}
\hat{A}=-A\phi+A^2+2\Omega^2+\frac{1}{4}m\rho\cos{\beta}+\frac{1}{4}\Omega\rho\cosh{\eta} \\
\hat{\Omega}=-\Omega\left(A+\phi\right) \\
\hat{\phi}= \!-\frac{1}{2}\phi^{2}-A\phi+2\Omega^2+\frac{1}{2}m\rho\cos{\beta} \\
\hat{\rho}=\rho\left(2m\cosh{\eta}\sin{\beta}+A-\phi\right)\\
\hat{\beta}=2m\cosh{\eta}\cos{\beta}-\Omega\\
\hat{\eta}=- 2m\sinh{\eta}\sin{\beta},
\end{cases}
\eea
describing a non-perfect spinorial fluid in an LRSI space-time. Defining the affine parameter along the spatial congruence by $x$, and indicating the boundary conditions as $f(x=0)=f_0$ with $f\in\left\{A,\Omega,\phi,\rho,\beta,\eta\right\}$, we have performed the numerical integration to derive the behavior of the spinorial, kinematic, and thermodynamic variables. The boundary conditions have been chosen over a wide range of possible values to explore the stability of the resulting numerical solutions. Even if not strictly necessary, here and in the following, we have chosen boundary values that satisfy the weak energy conditions. In this way, one can assume that the effective fluid obtained is as close as possible to the standard picture of incoherent matter\footnote{The more exotic phenomenology associated with the fermion field will be investigated with more powerful methods in future works.} and check the stability of such a feature throughout the numerical solutions presented. We have experimented with a wide range of boundary conditions that satisfy the weak energy condition. An example of the results obtained is shown in Figs.~\ref{Fig:LRSI1} and~\ref{Fig:LRSI2}.  All quantities are presented in dimensionless form, derived by normalizing each variable with respect to the mass $m$ and its associated Compton length.


Numerical solutions displayed in Figs.~\ref{Fig:LRSI1} and~\ref{Fig:LRSI2} appear to be stable under several different sets of boundary conditions and reveal several noteworthy features. 
The relationship \eqref{eosLRSInp} between the energy density $\mu$ and the tangential pressure $p_o$ of the effective fermionic fluid suggests a matter distribution composed of stiff shells. 
The acceleration $A$, interpreted as the surface gravity of the configuration, exhibits a nontrivial sign change approximately halfway to the surface. 
Moreover, the quantity $\phi$, associated with the (pseudo-)Gaussian curvature of the bispaces, vanishes at a finite value of the affine parameter $x$. 
Both features warrant further investigation and will be examined in future works.  
Figure~\ref{Fig:LRSI2} illustrates the behavior of the quantities associated with the spinorial fluid: the chiral angle $\beta$ approaches an asymptotically constant value, as does the parameter $\eta$. 
In particular, $\eta$ tends to zero, indicating that, as $x$ increases, the 4-velocity vector field $u_a$ and the spin axial vector field $s_a$ progressively align with the unit vectors $v_a$ and $e_a$ tangent to the congruences.

A particularly relevant aspect (see Fig.~\ref{Fig:LRSI1}) concerns the behavior of the radial pressure $p_r$, which vanishes at a finite value $x_0$ of the parameter $x$. By contrast, the vorticity remains regular throughout and is consistently negative, attaining a maximum near this point $x_0$. In the case of a standard fluid, the fact that $p_r$ vanishes at a finite point would allow one, via Israel's junction conditions, to interpret the solution up to $x_0$ as describing the interior of a compact matter distribution. Since the weak energy condition is preserved (see Fig. \ref{Fig:LRSI1}), this result suggests an intriguing conclusion: in this configuration, the spinor field might be used as a semiclassical model for a relativistic star. This would necessarily be a toy model, analogous to that used for boson stars, in which the matter source is a classical scalar field (see e.g. \cite{Schunck:2003kk}). Yet, at present, in general relativity, no known vortical model for the interior of a relativistic star exists (in our knowledge), and the solution we found could be a good occasion to explore more deeply the properties of such solutions.   Another interesting interpretation of the solution considered above would be a {\it classical} fluid representation of spin-1/2 particles, e.g., the electron or the proton. This is consistent with experimental studies \cite{BEG,Polyakov:2018zvc,Lorce:2025oot,Lorce:2018egm} that measured effective isotropic and shear pressures within a proton. 

While certainly intriguing, these ideas are not free of caveats. For example, the two interpretations above require a detailed analysis of the validity of explicit junction conditions for the Dirac field equations, and one should also verify that, in the full quantum regime, this solution remains valid. These two tasks and other issues that might arise warrant further investigation, which is left for future work.

\subsubsection{Perfect spinorial fluid in homogeneous anisotropic LRSII space-times.}
We consider the system \eqref{fieleqphi0homo} which describes a spinorial dust in homogeneous anisotropic LRSII space-times
\begin{equation}\label{fieleqphi0homo_}
\begin{cases}
&\dot{\Theta}=-\frac{1}{3}\Theta^{2}-\frac{3}{2}\Sigma^{2}-\frac{1}{4}m\rho\cos\beta \\
&\dot{\Sigma}=\frac{1}{2}\Sigma^{2}-\frac{2}{3}\Theta\Sigma-\left(\Sigma-\frac{1}{3}\Theta\right)\left(\Sigma+\frac{2}{3}\Theta\right)-\frac{1}{3}m\rho\cos\beta \\
&\dot{\rho}=\rho\left(-\Theta+2m\sinh{\eta}\sin{\beta}\right) \\
&\dot{\eta}=-2m\cosh{\eta}\sin{\beta} \\
&\dot{\beta}=2m\cos{\beta}\sinh{\eta} .
\end{cases}
\end{equation}
The energy density $\mu$ and the electric scalar part $E$ of the Weyl tensor are given in \eqref{energyLRSII} and \eqref{constraint_phi0}, respectively. 
After the introduction of the affine parameter $t$ along the time-like congruence and setting the mass of the spinor field $m = 1$, we choose initial values for the unknown $\left\{\Omega,\Sigma,\rho,\beta,\eta\right\}$, 
given by $\Theta(t=1.6)=\rho(t=1.6)=\eta(t=1.6)=1$, $\Sigma(t=1.6)=10^{-5}$ and $\beta(t=1.6)=0$.

\begin{figure}[H]
    \centering
    \begin{minipage}[b]{0.45\textwidth}
        \centering
        \includegraphics[height=4.8cm]{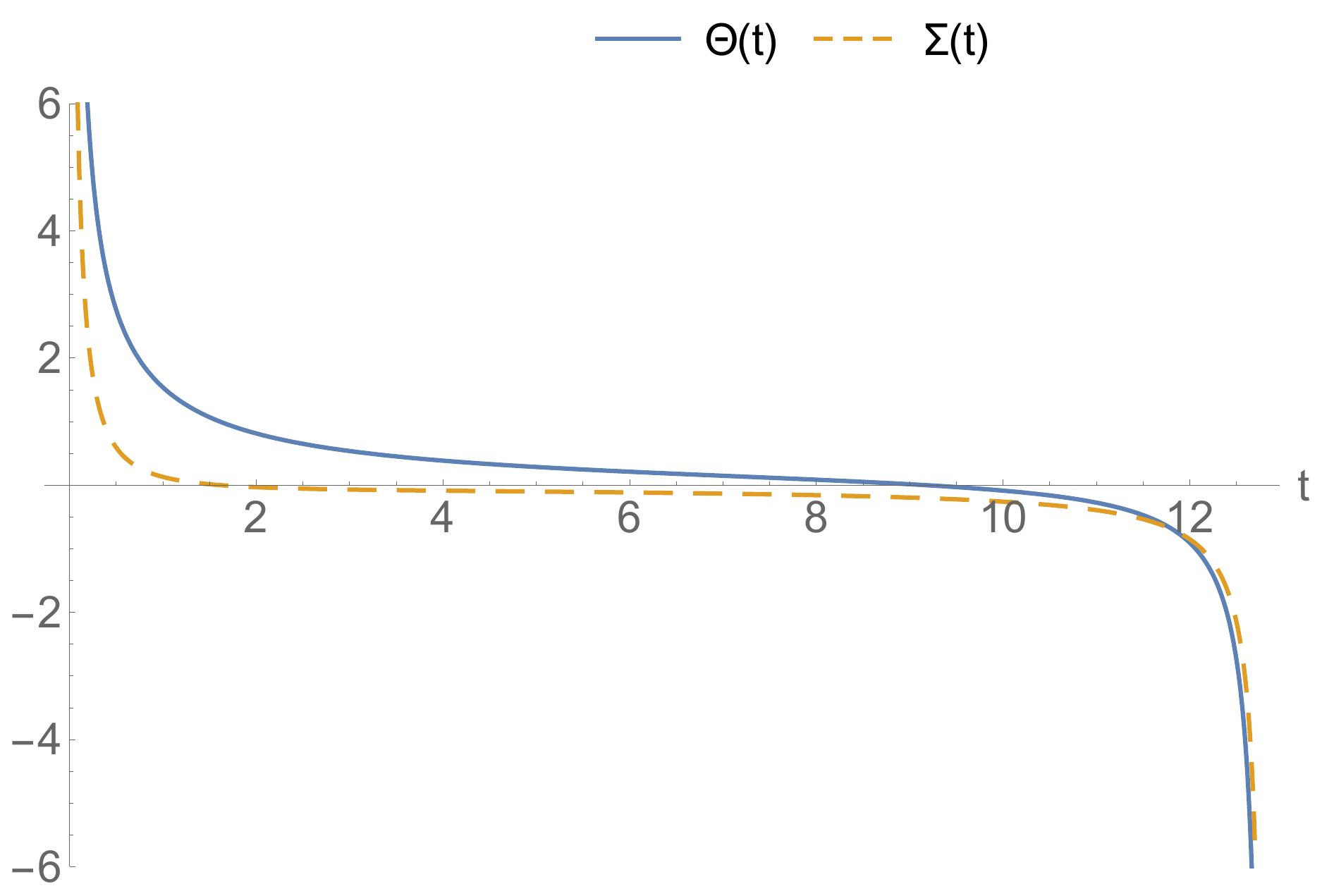}
    \end{minipage}
    \hspace{0.05\textwidth}
    \begin{minipage}[b]{0.45\textwidth}
        \centering
        \includegraphics[height=4.8cm]{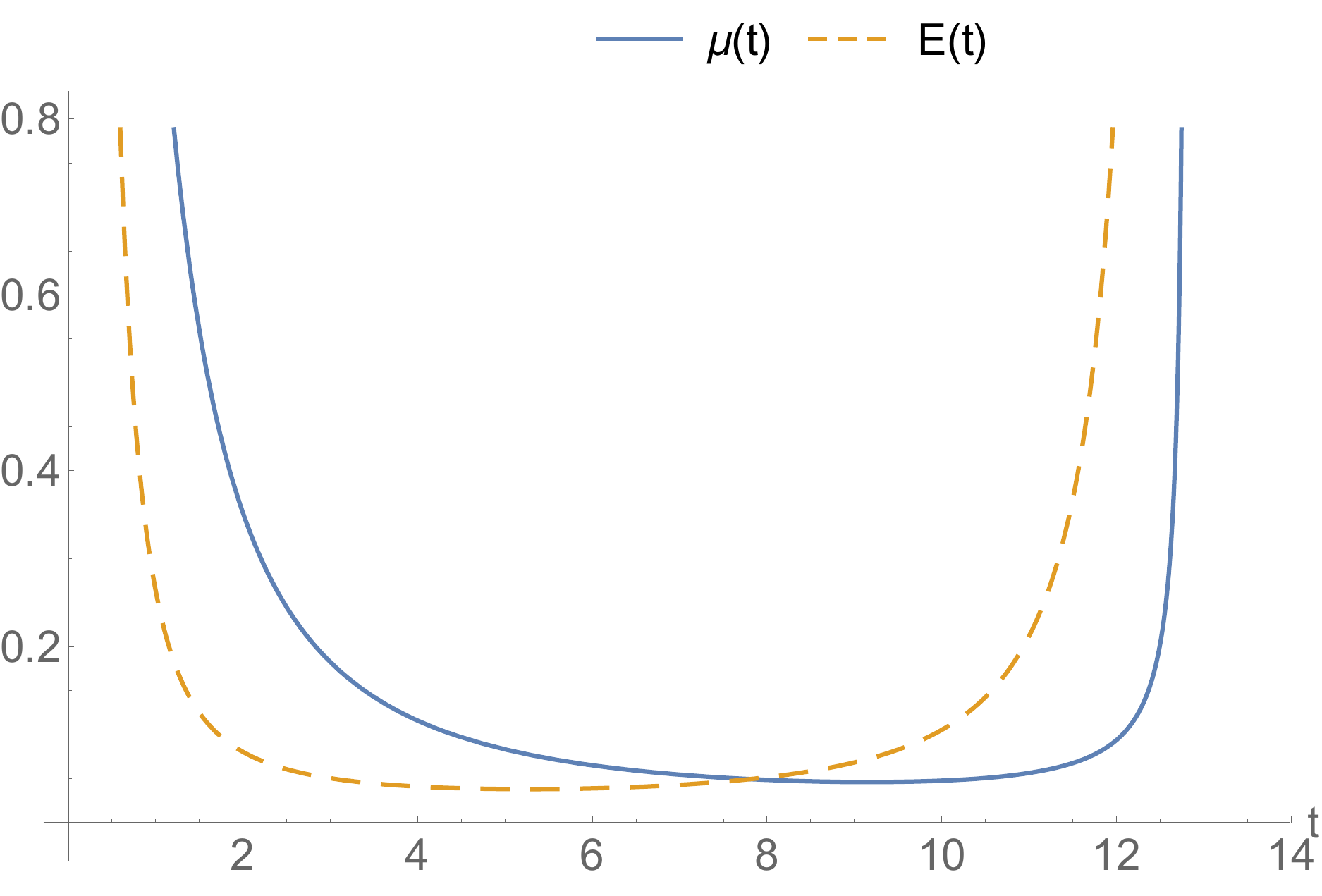}
    \end{minipage}
\caption{\small Dynamical evolution of the system \eqref{fieleqphi0homo_} with $m = 1$ and initial data $\Theta(t=1.6)=\rho(t=1.6)=\eta(t=1.6)=1$, $\Sigma(t=1.6)=10^{-5}$, $\beta(t=1.6)=0$. Left-hand panel: behavior of the kinematical variables $\Theta$ (solid line), $\Sigma$ (dotted line) in function of the affine parameter $t$. Right-hand panel: behavior of the energy density $\mu$ (solid line) and the electric part $E$ (dashed line) of the Weyl tensor as functions of $t$.}
\label{Fig:BianchiI1}
\end{figure}
  \begin{figure}[H]
 \centering
    \begin{minipage}[b]{0.45\textwidth}
        \centering
        \includegraphics[height=4.8cm]{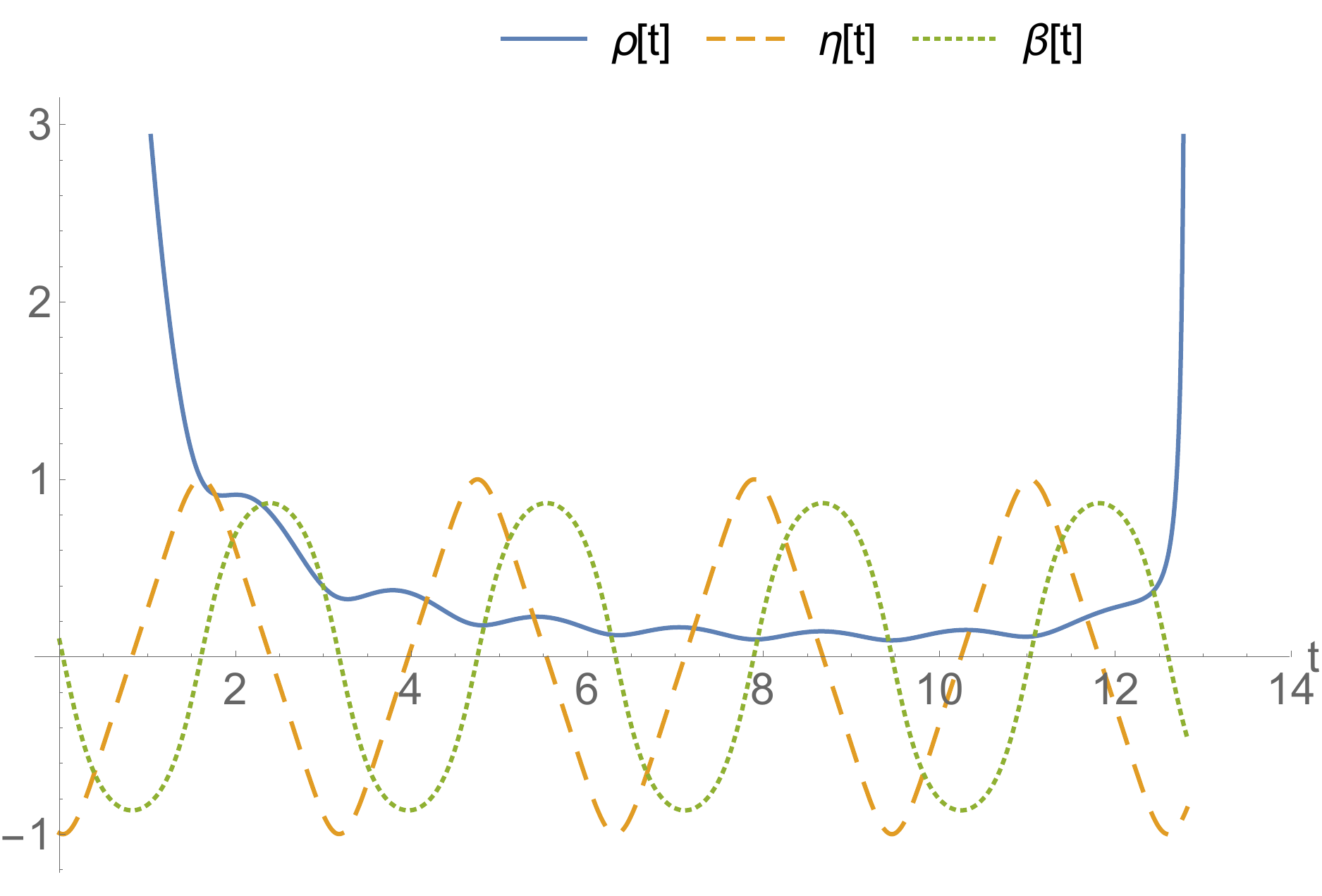}
 \end{minipage}
  \hspace{0.05\textwidth}
    \begin{minipage}[b]{0.45\textwidth}
        \centering
        \includegraphics[height=4.8cm]{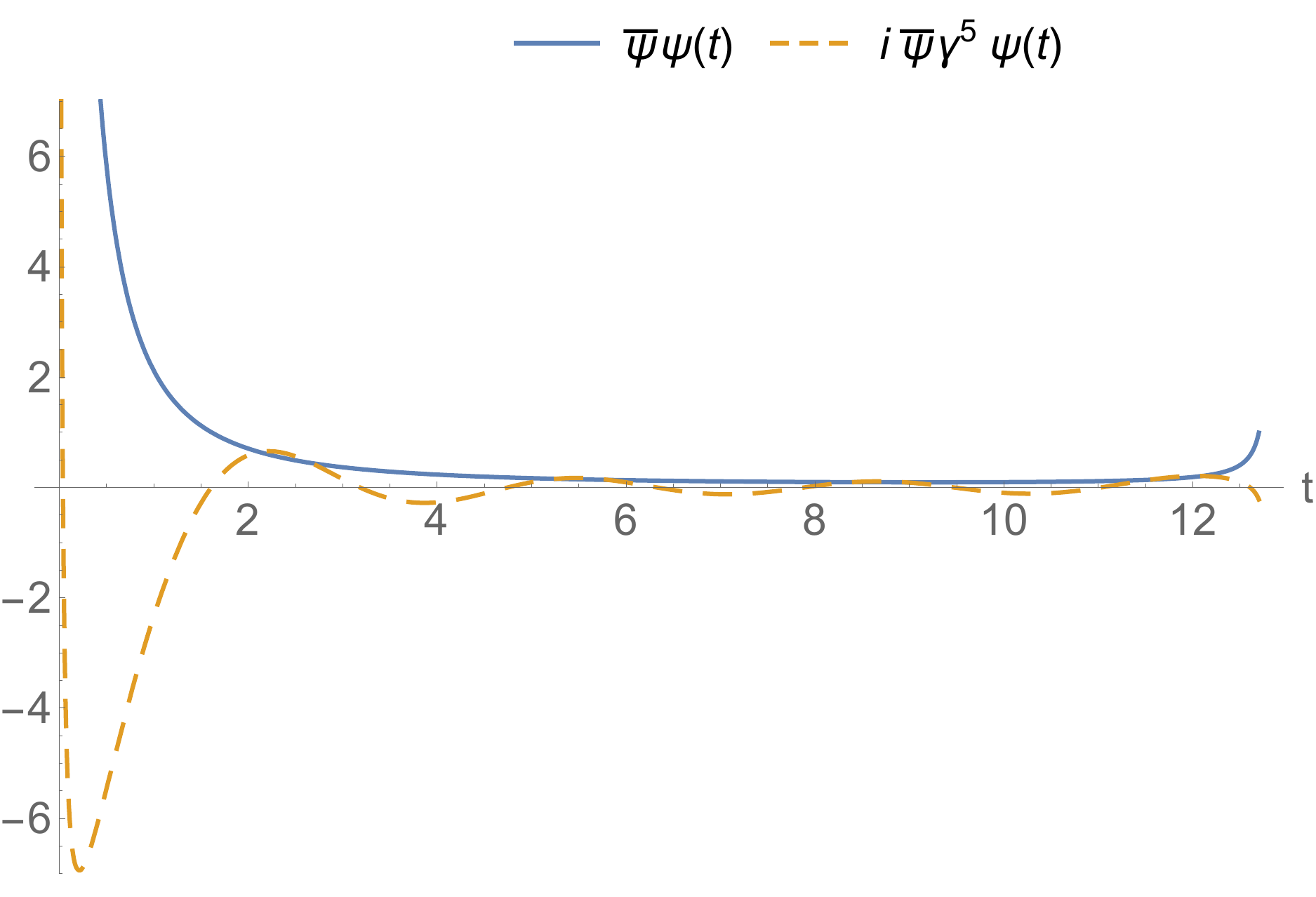}
 \end{minipage}
        \caption{\small Dynamical evolution of the system \eqref{fieleqphi0homo_} with $m = 1$ and initial data $\Theta(t=1.6)=\rho(t=1.6)=\eta(t=1.6)=1$, $\Sigma(t=1.6)=10^{-5}$, $\beta(t=1.6)=0$. Left-hand panel: dynamical behavior of the modulus $\rho$ (solid line), the pseudo-scalar function $\eta$ (dashed line), and the chiral angle $\beta$ (dotted line) in function of the affine parameter $t$. Right-hand panel: behavior of the scalar $\bar{\psi}\psi$ (solid line) and the pseudo-scalar $i\bar{\psi}\gamma^5\psi$ (dashed line) in function of $t$.}
\label{Fig:BianchiI2}
\end{figure}
\begin{figure}[H]
\centering
        \includegraphics[height=5.0cm]{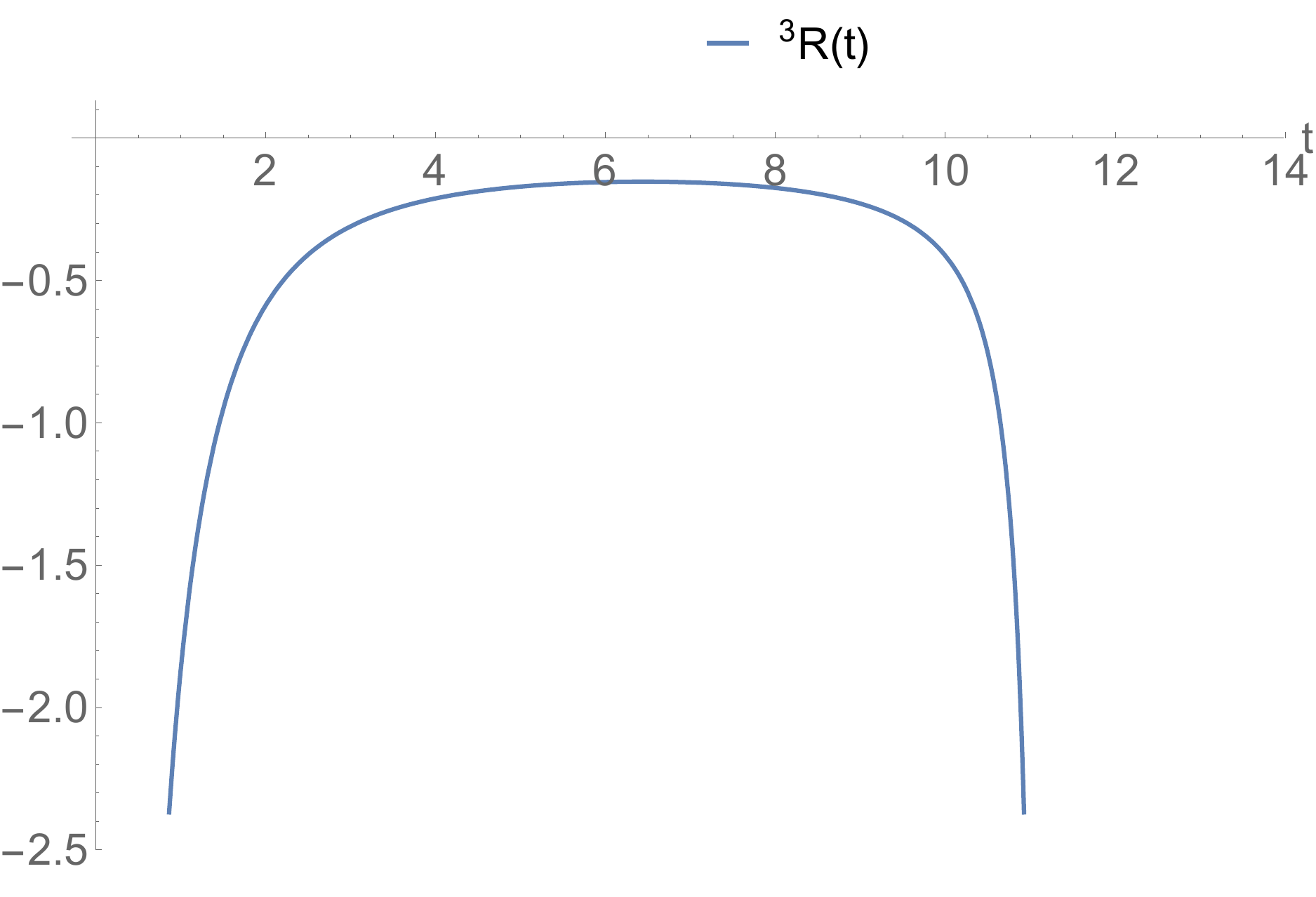}
\caption{\small The behavior of the $3$-Ricci curvature with initial data $m = 1$, $\Theta(t=1.6)=\rho(t=1.6)=\eta(t=1.6)=1$, $\Sigma(t=1.6)=10^{-5}$ and $\beta(t=1.6)=0$.}
\label{BianchiFS_1}
\end{figure}
\begin{figure}[H]
    \centering
    \begin{minipage}[b]{0.45\textwidth}
        \centering
        \includegraphics[height=4.8cm]{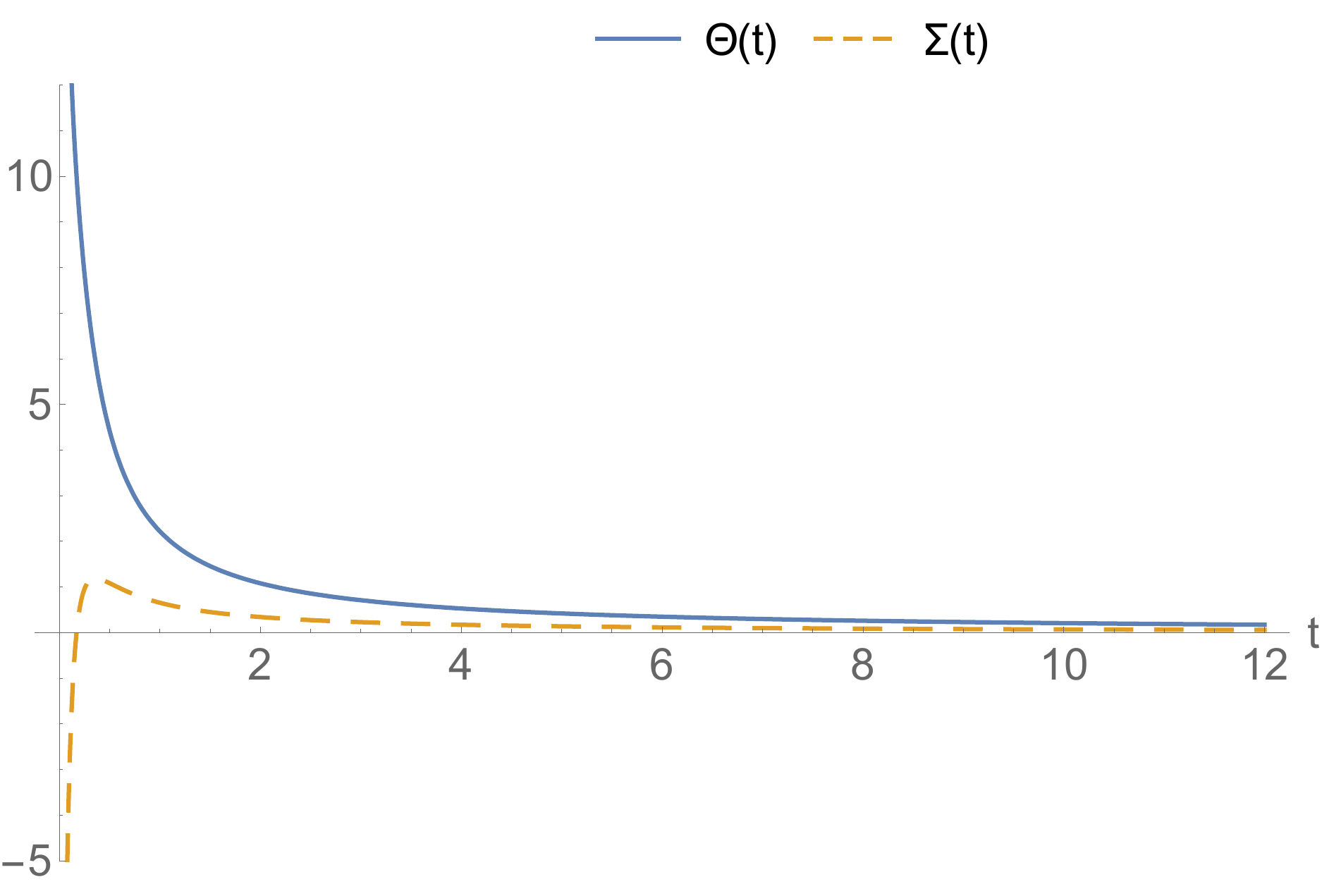}
    \end{minipage}
    \hspace{0.05\textwidth}
    \begin{minipage}[b]{0.45\textwidth}
        \centering
        \includegraphics[height=4.8cm]{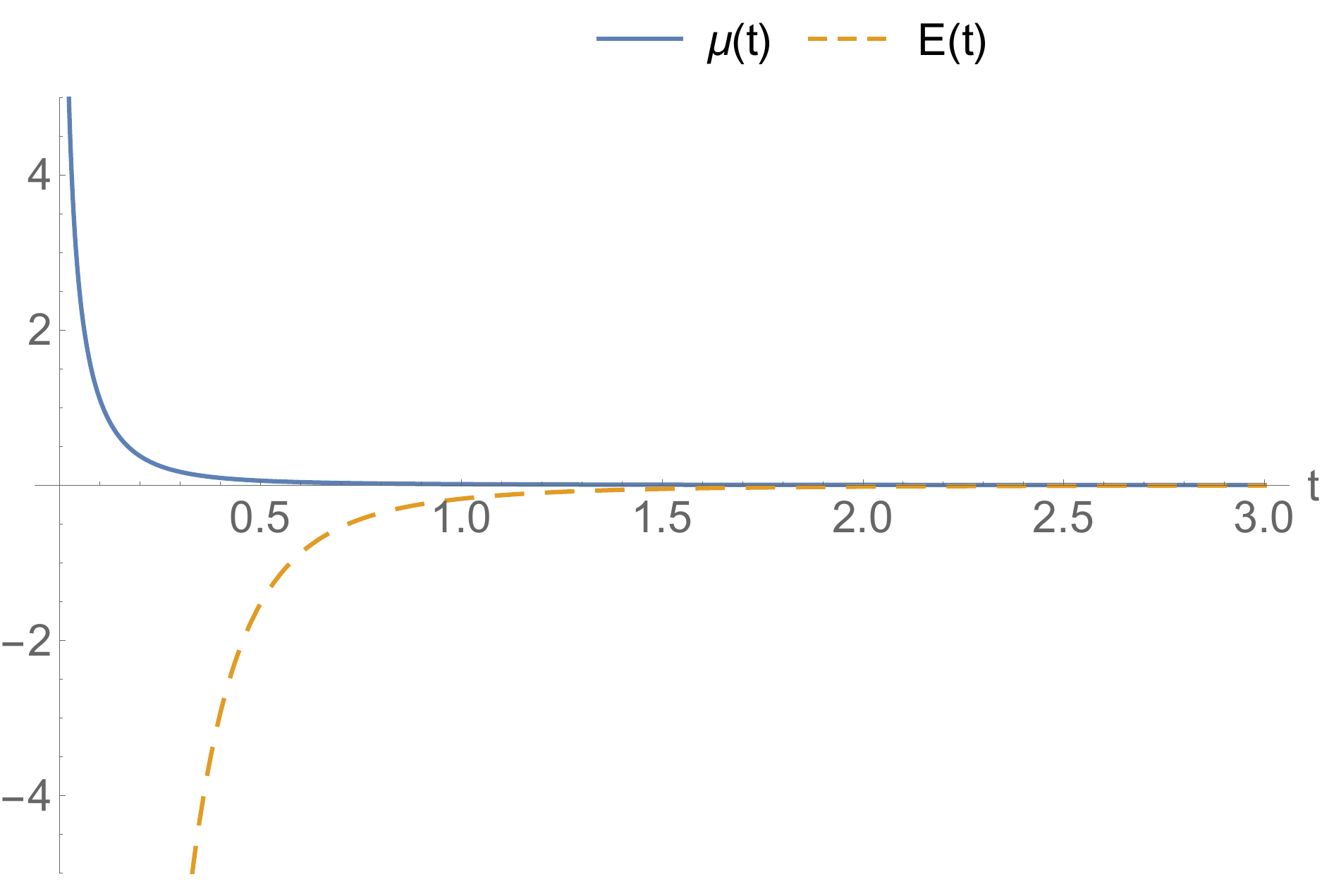}
    \end{minipage}
\caption{\small Dynamical evolution of the system \eqref{fieleqphi0homo_} with $m = 1$ and initial data $\Theta(t=0.17)=10$, $\rho(t=0.17)=\eta(t=0.17)=1$, $\Sigma(t=0.17)=10^{-5}$ and $\beta(t=0.17)=0$. Left-hand panel: behavior of the kinematical variables $\Theta$ (solid line), $\Sigma$ (dotted line) in function of the affine parameter $t$. Right-hand panel: behavior of the energy density $\mu$ (solid line) and the electric part $E$ (dashed line) of the Weyl tensor in function of the affine parameter $t$.}
\label{Fig:BianchiI3}
   \end{figure}
\begin{figure}[H]
 \centering
    \begin{minipage}[b]{0.45\textwidth}
        \centering
        \includegraphics[height=4.8cm]{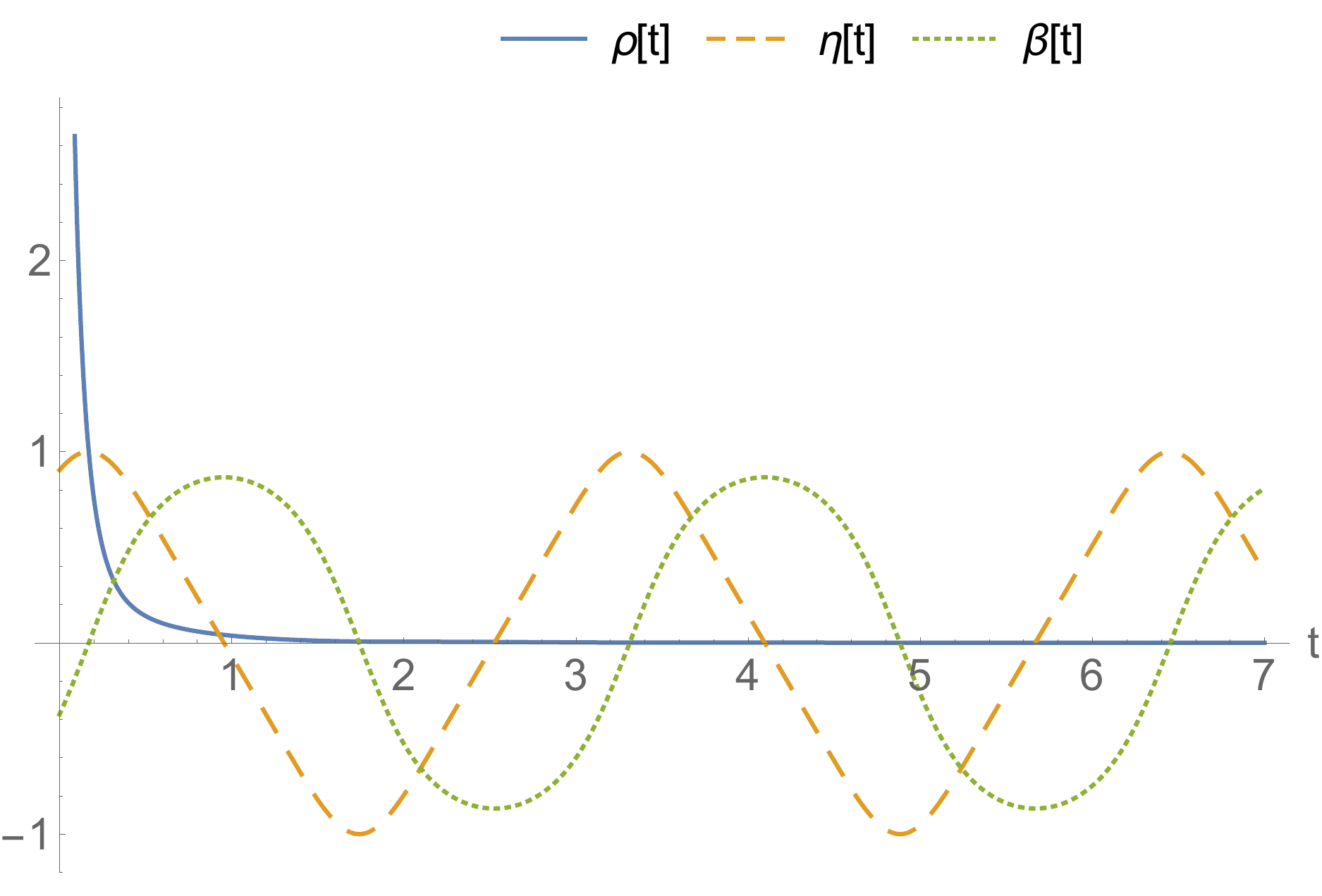}
 \end{minipage}
  \hspace{0.05\textwidth}
    \begin{minipage}[b]{0.45\textwidth}
        \centering
        \includegraphics[height=4.8cm]{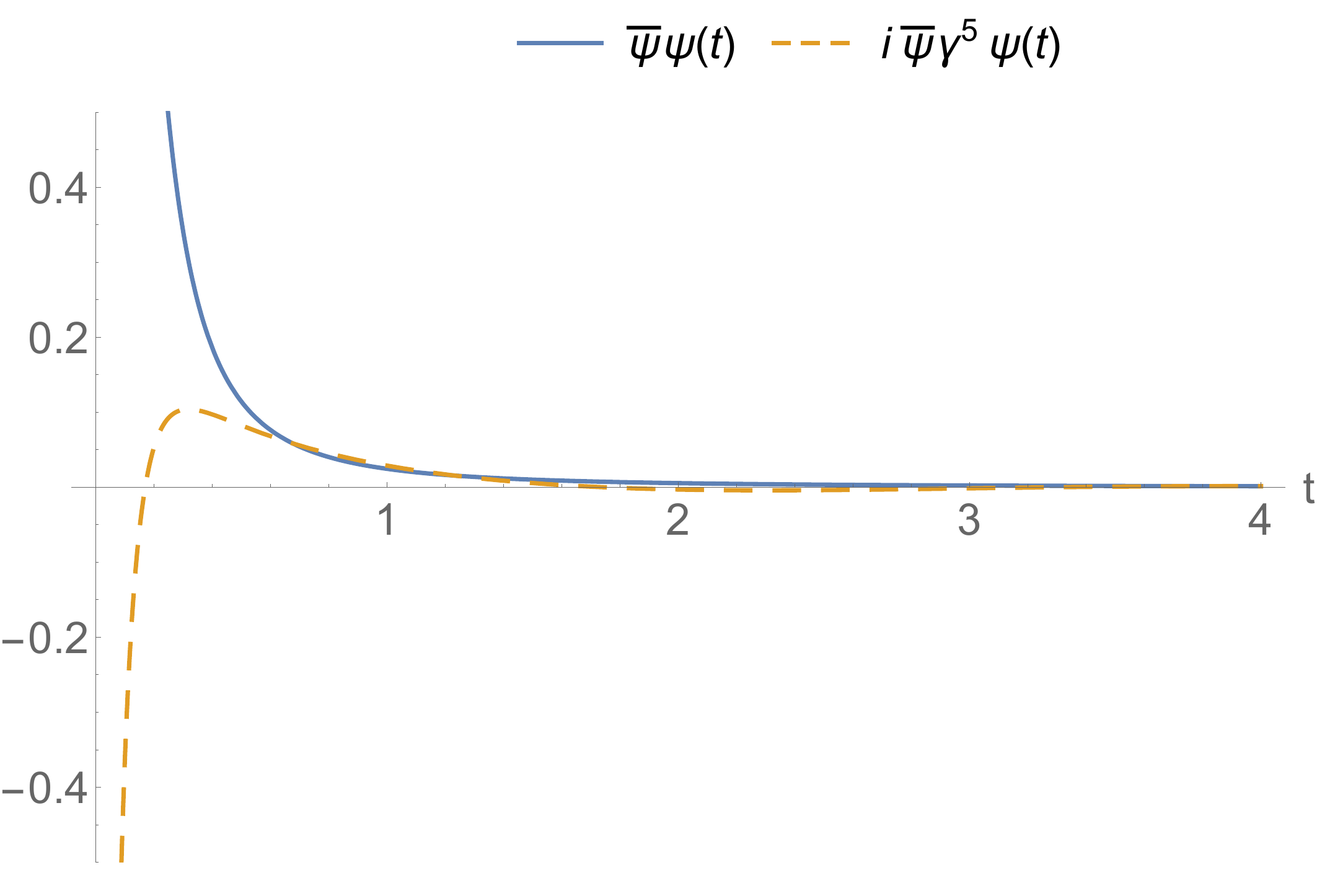}
 \end{minipage}
        \caption{\small Dynamical evolution of the system \eqref{fieleqphi0homo_} with $m = 1$ and initial data $\Theta(t=0.17)=10$, $\rho(t=0.17)=\eta(t=0.17)=1$, $\Sigma(t=0.17)=10^{-5}$, $\beta(t=0.17)=0$. Left-hand panel: dynamical behavior of the modulus $\rho$ (solid line), the pseudo-scalar function $\eta$ (dashed line), and the chiral angle $\beta$ (dotted line) as functions of the affine parameter $t$. Right-hand panel: behavior of the scalar $\bar{\psi}\psi$ (solid line) and the pseudo-scalar $i\bar{\psi}\gamma^5\psi$ (dashed line) in function of the affine parameter $t$.}
\label{Fig:BianchiI4}
\end{figure}

To better illustrate the sensitivity of system \eqref{fieleqphi0homo_} with respect to the initial expansion rate, we consider a second set of initial conditions. While all parameter and initial variable values are kept identical to the previous case, the initial time is shifted from $t=1.6$ to $t=0.17$ with $\Theta(t=0.17)=10$, allowing for a clearer dynamic visualization of the numerical solution.
The numerical integration with this modified initial condition is shown in Fig.~\ref{Fig:BianchiI3} and \ref{Fig:BianchiI4}.
\begin{figure}[H]
    \centering
\includegraphics[height=5.0cm]{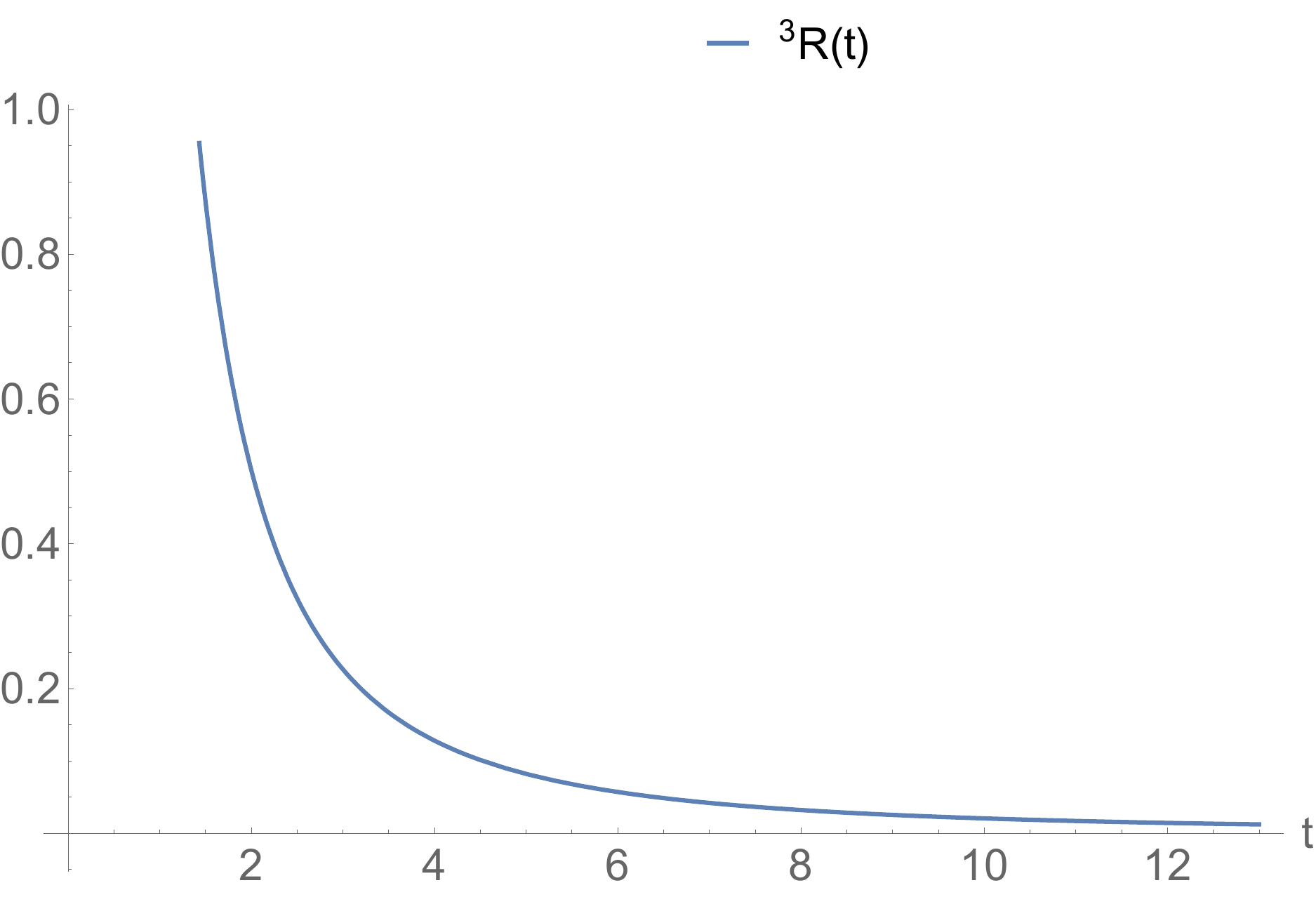}
\caption{\small  The behavior of the $3$-Ricci curvature with initial data $m = 1$, $\Theta(t=0.17)=10$, $\rho(t=0.17)=\eta(t=0.17)=1$, $\Sigma(t=0.17)=10^{-5}$ and $\beta(t=0.17)=0$.}
\label{BianchiFS_2}
\end{figure}

The figures above illustrate the two qualitatively distinct dynamical regimes which appear to be stable under changes of initial conditions.
In both cases, the evolution originates from an initial singularity, where the expansion rate $\Theta$ diverges positively. The shear $\Sigma$ diverges at early times, signaling an anisotropic initial singularity. 
The first scenario can be interpreted as describing an anisotropic universe that undergoes an initial phase of expansion followed by a gravitational collapse. 
The behavior of the shear scalar $\Sigma$ indicates that the final singularity retains a pronounced anisotropic character. 
In the second case, no collapse occurs, and the space-time exhibits a progressive isotropization as the expansion proceeds. 

To clarify the physical understanding of these solutions, it is also useful to consider the 3-Ricci scalar 
\begin{equation}\label{Friedmann_constraint}
    {}^{3}R=-2\mu+\tfrac23\Theta^2-\tfrac32\Sigma^2,
\end{equation}
obtained by the Gauss embedding equation and the Ricci identities for $v^a$.  In Figure~\ref{BianchiFS_1} ${}^{3}R$ is negative, so the collapsing model has a closed geometry, connecting it to a Kantowski-Sachs model. In Figure~\ref{BianchiFS_2} ${}^{3}R$ is positive, indicating an open spatial geometry typical of the diagonal Bianchi III models. This result is consistent with the well-known properties of the homogeneous and isotropic case: closed spatial geometries tend to collapse, whereas open geometries are associated with an unlimited expansion.

As already observed in the homogeneous and isotropic solution discussed in the previous Subsection, the dynamics of the spinor field only partially affect the evolution of the space-time, which is primarily governed by the effective thermodynamical quantities.

\subsubsection{Non-perfect spinorial fluid in LRSIII space-times}
We now turn to the system \eqref{eq_finali_LRSIII_np}
\begin{eqnarray}\label{eq_finali_LRSIII_np_}
\begin{cases}
&\dot{\xi}=-\dfrac{1}{3}\xi\left(\Theta+6\Sigma\right)\\[6pt]
&\dot{\Theta}=-\dfrac{1}{3}\Theta^2-\dfrac{3}{2}\Sigma^2-\dfrac{1}{4}m\rho\cos{\beta}-\frac{1}{4}\rho\xi\sinh{\eta}\\[6pt]
&\dot{\Sigma}=-\dfrac{1}{2}\Sigma^2-\Theta\Sigma+\dfrac{2}{9}\Theta^2+2\xi^2-\dfrac{1}{3}m\rho\cos{\beta}+\dfrac{1}{6}\rho\xi\sinh{\eta}\\[6pt]
&\dot{\rho}=\rho\left(2m\sinh{\eta}\sin{\beta}-\Theta\right)\\[6pt]
&\dot{\eta}=-2m\cosh{\eta}\sin{\beta}\\[6pt]
&\dot{\beta}=2m\sinh{\eta}\cos{\beta}-\xi,
\end{cases}
\end{eqnarray}
where the thermodynamic quantities are expressed as in eqs. \eqref{thermo_LRSIII_np}. Once again, the evolution is parametrized by the affine parameter $t$, with initial conditions given for $\{\xi,\Sigma,\Theta,\rho,\eta,\beta\}$. We set $m=1$ and initial values $\xi(t=0.12)=7$, $\Theta(t=0.12)=10$, $\rho(t=0.12)=\eta(t=0.12)=1$, $\Sigma(t=0.12)=10^{-5}$, $\beta(t=0.12)=\tfrac\pi 2$, in order to satisfy, also in this case, the constraints imposed by the weak energy condition.
\begin{figure}[H]
    \centering
    \begin{minipage}[b]{0.45\textwidth}
        \centering
        \includegraphics[height=5.0cm]{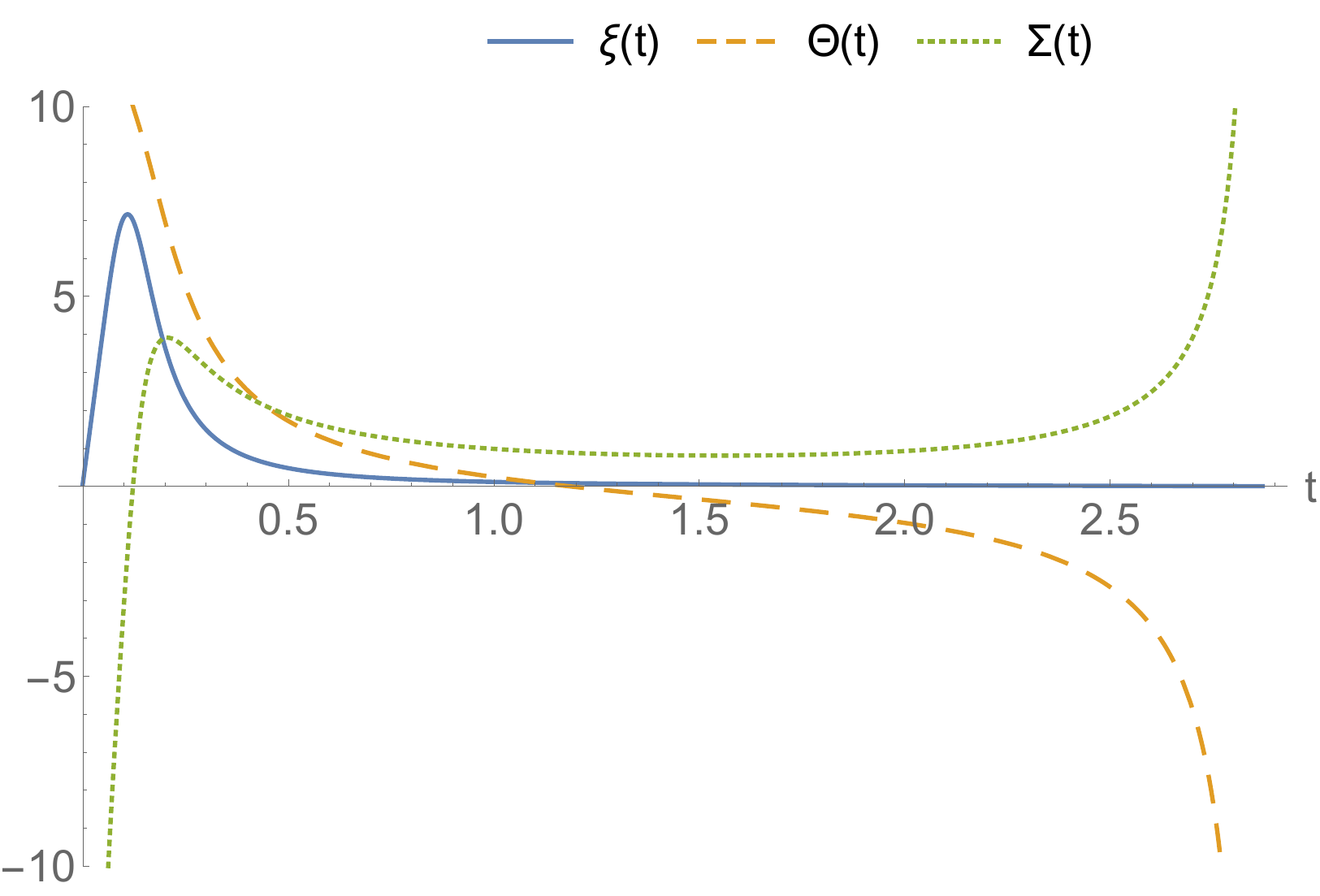}
    \end{minipage}
    \hspace{0.05\textwidth}
    \begin{minipage}[b]{0.45\textwidth}
        \centering
        \includegraphics[height=5.0cm]{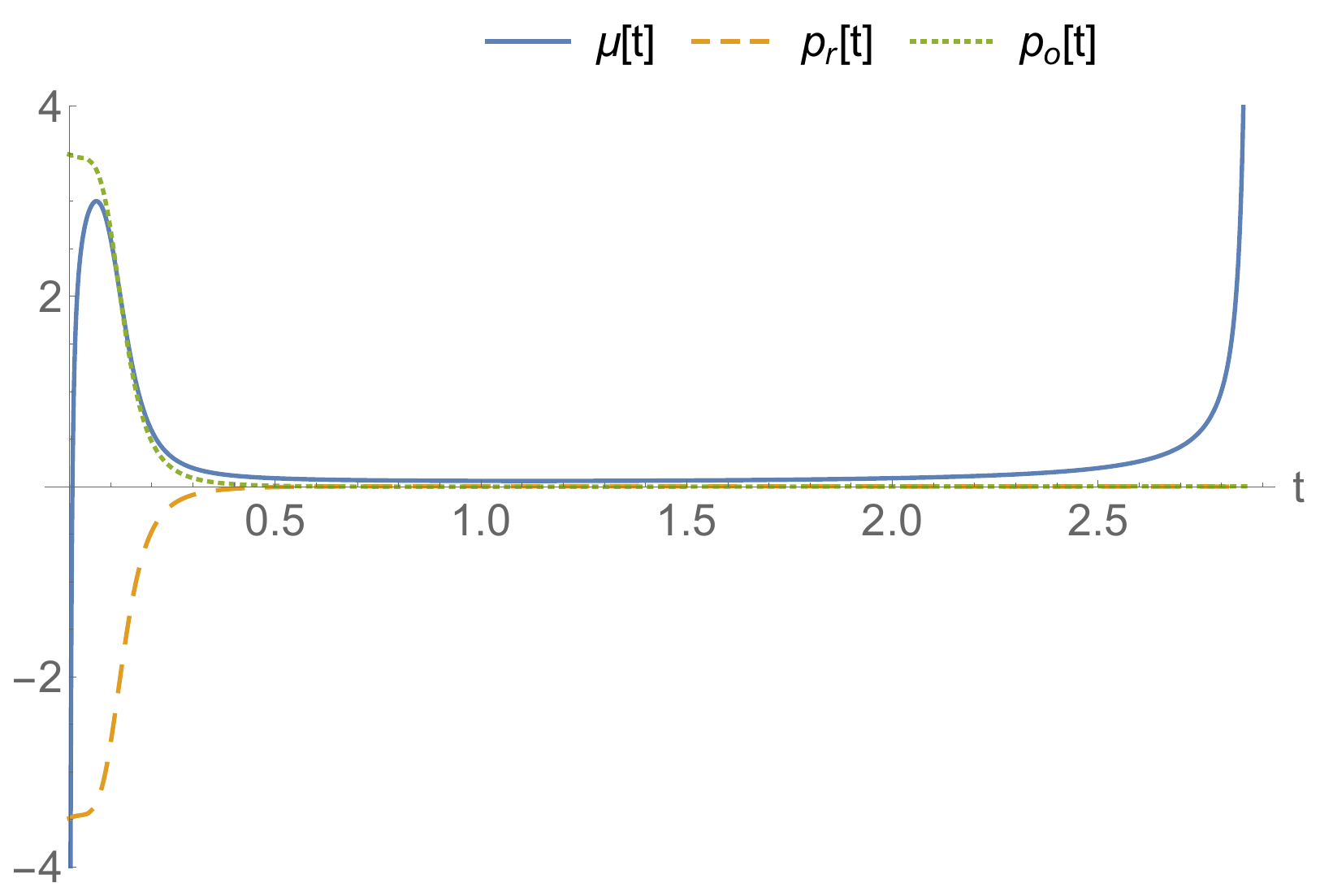}
    \end{minipage}
\caption{\small Dynamical evolution of the system \eqref{eq_finali_LRSIII_np_} with $m = 1$ and initial data $\xi(t=0.12)=7$, $\Theta(t=0.12)=10$, $\rho(t=0.12)=\eta(t=0.12)=1$, $\Sigma(t=0.12)=10^{-5}$, $\beta(t=0.12)=\tfrac\pi 2$. Left-hand panel: behavior of the kinematical variables $\xi$ (solid line), $\Sigma$ (dotted line), and $\Theta$ (dashed line) as functions of the affine parameter $t$. Right-hand panel: behavior of the energy density $\mu$ (solid line), radial pressure $p_r$ (dashed line) and orthogonal pressure $p_o$ (dotted line) in function of $t$.}
\label{Fig:LRSIII1}
\end{figure}
\begin{figure}[H]
 \centering
    \begin{minipage}[b]{0.45\textwidth}
        \centering
        \includegraphics[height=5.0cm]{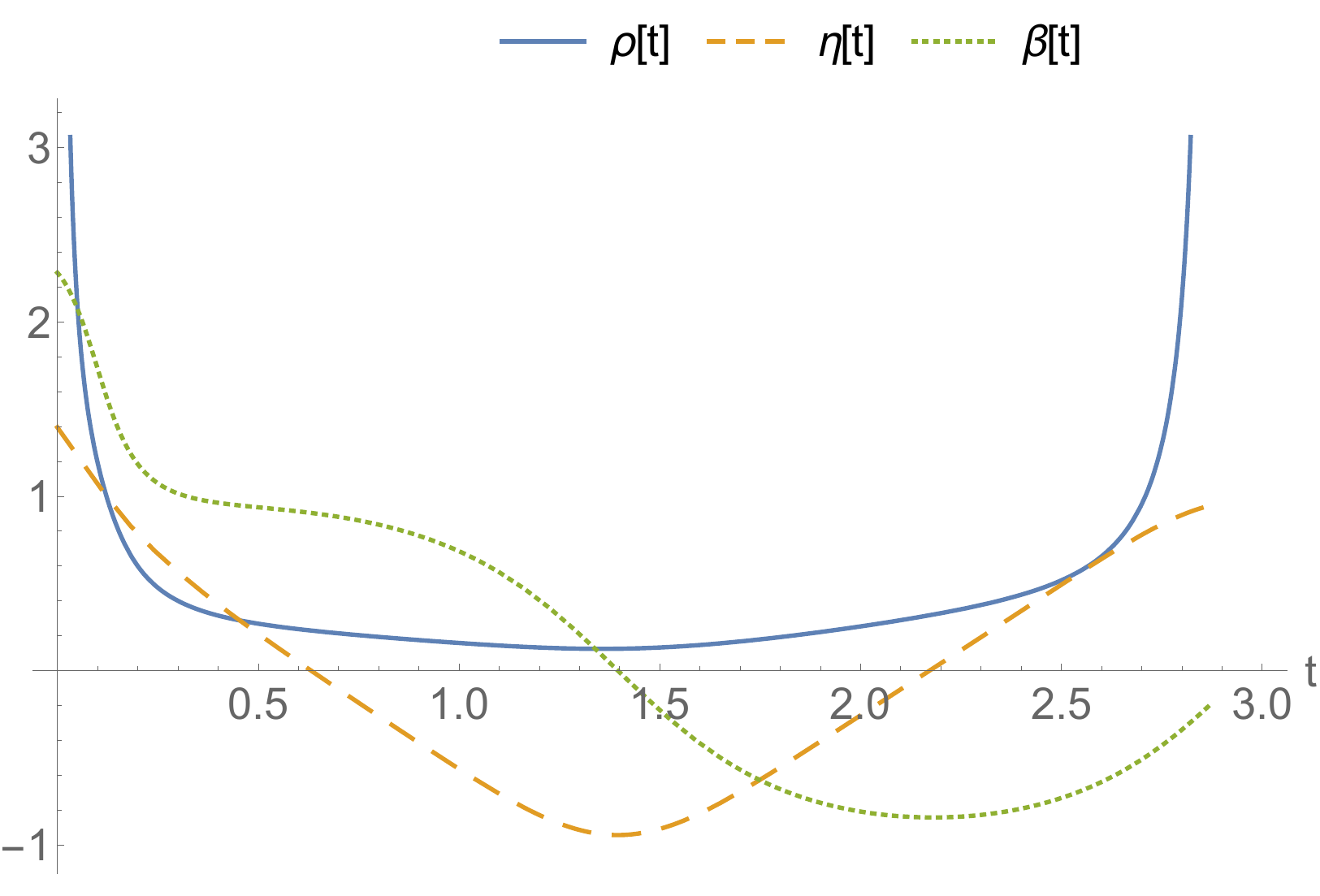}
 \end{minipage}
  \hspace{0.05\textwidth}
    \begin{minipage}[b]{0.45\textwidth}
        \centering
        \includegraphics[height=5.0cm]{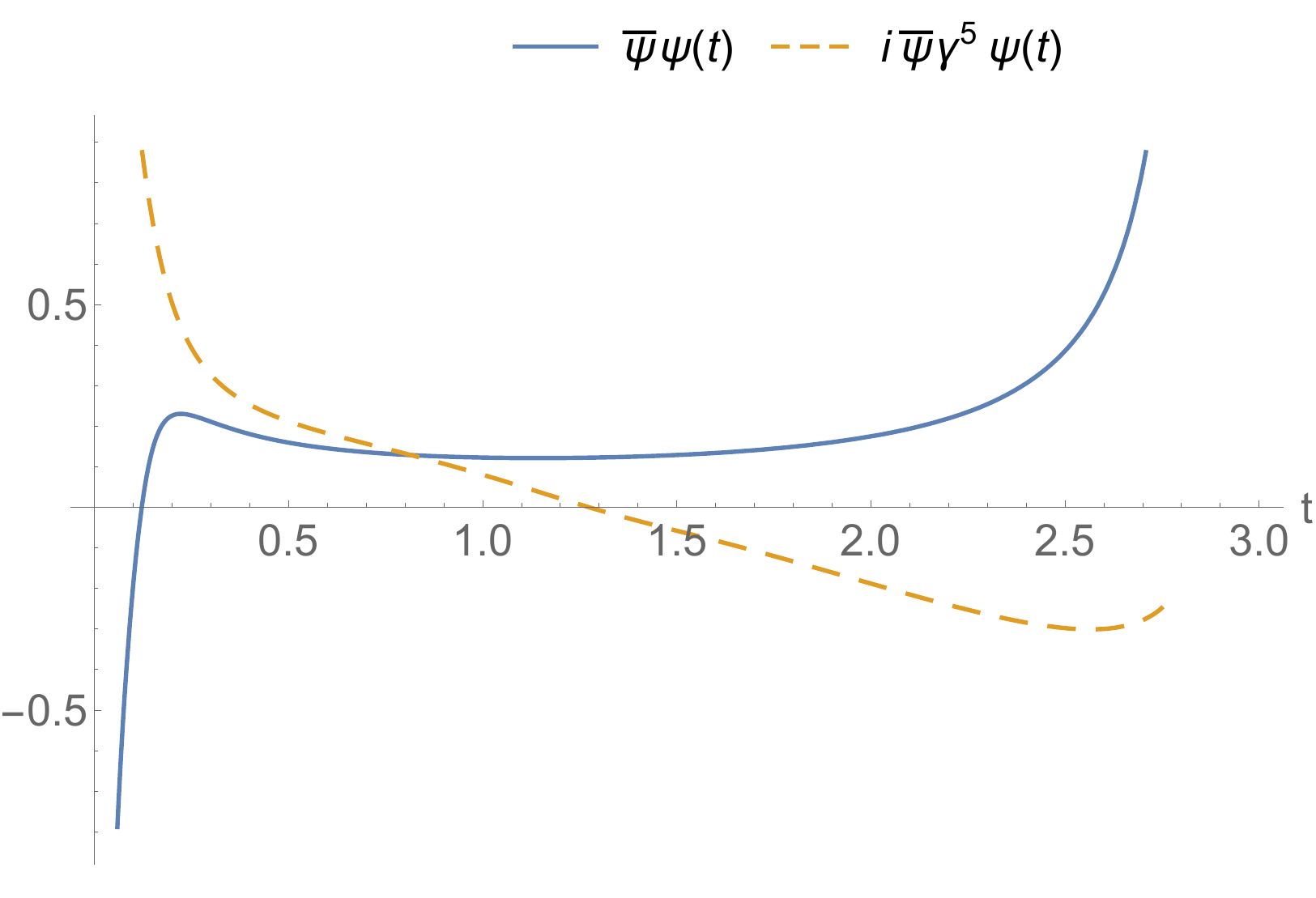}
 \end{minipage}
        \caption{\small Dynamical evolution of the system \eqref{eq_finali_LRSIII_np} with $m = 1$ and initial data $\xi(t=0.12)=7$, $\Theta(t=0.12)=10$, $\rho(t=0.12)=\eta(t=0.12)=1$, $\Sigma(t=0.12)=10^{-5}$, $\beta(t=0.12)=\tfrac\pi 2$. Left-hand panel: dynamical behavior of the modulus $\rho$ (solid line), the pseudo-scalar function $\eta$ (dashed line) and the chiral angle $\beta$ (dotted line) in function of the affine parameter $t$. Right-hand panel: behavior of the scalar $\bar{\psi}\psi$ (solid line) and the pseudo-scalar $i\bar{\psi}\gamma^5\psi$ (dashed line) in function of $t$.}
\label{Fig:LRSIII2}
\end{figure}
\begin{figure}[H]\label{LRSIII_FS}
\centering
\includegraphics[height=5.0cm]{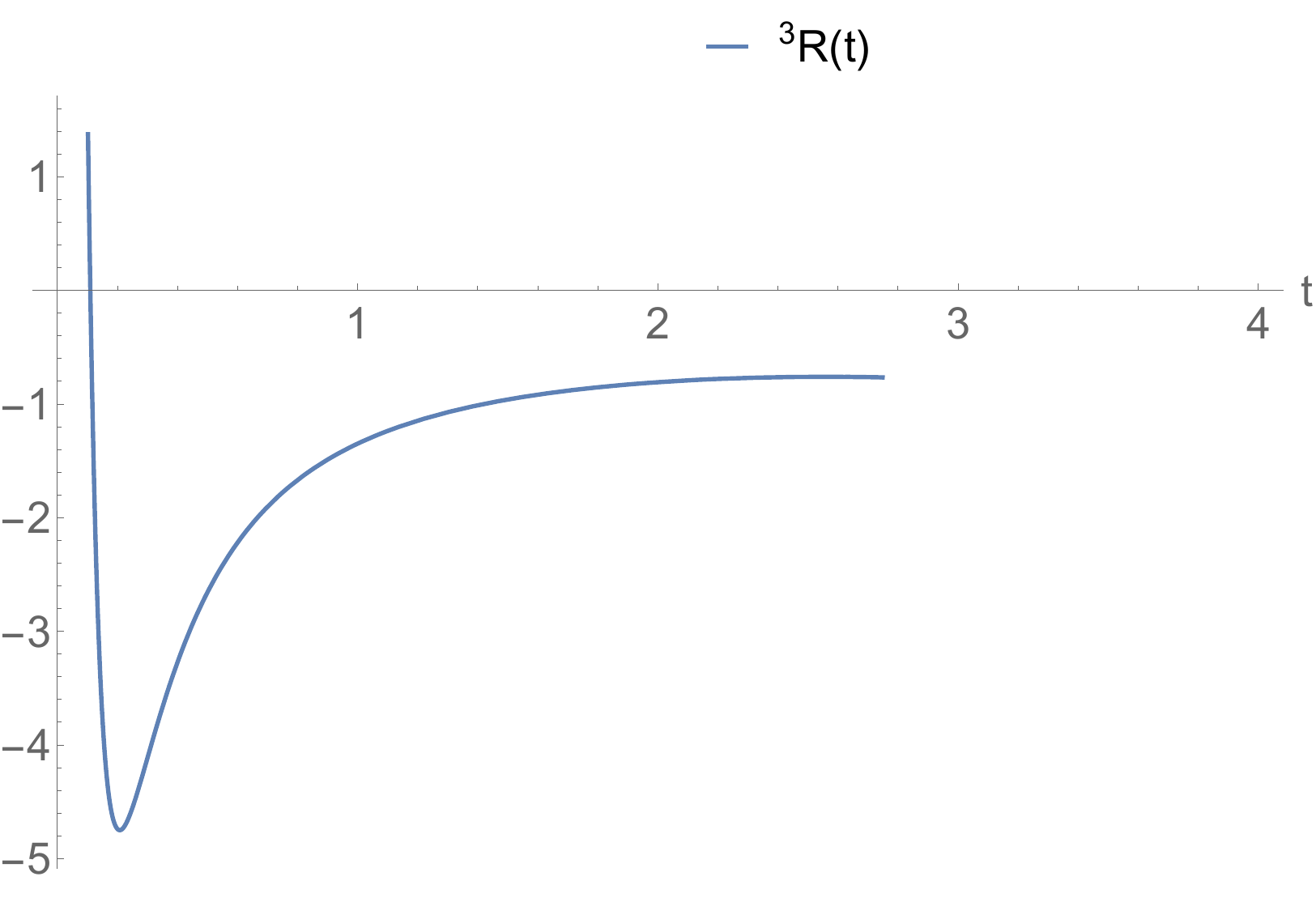}
\caption{\small The behavior of the $3$-Ricci scalar, with initial data $m = 1$, $\xi(t=0.12)=7$, $\Theta(t=0.12)=10$, $\rho(t=0.12)=\eta(t=0.12)=1$, $\Sigma(t=0.12)=10^{-5}$ and $\beta(t=0.12)=\tfrac\pi 2$.}
\end{figure}

The dynamical evolution of the kinematical variables $\{\Theta, \Sigma, \xi\}$, the thermodynamic quantities $\{\mu, p_r, p_o\}$, and the spinorial field variables $\{\rho, \beta, \eta\}$ is illustrated in Figs.~\ref{Fig:LRSIII1} and \ref{Fig:LRSIII2} in terms of the affine parameter $t$. These plots appear to be stable under changes of initial conditions, and describe a homogeneous space-time of Bianchi type II, VIII, or IX that, despite an initial non-vanishing twist, rapidly evolves toward a collapsing, spatially closed Kantowski-Sachs geometry delimited by two anisotropic gravitational singularities. Such geometry might at first sight appear different from the collapsing solutions found in the previous Subsection in that the shear scalar diverges positively rather than negatively. However, this difference is only coordinate-related, as it becomes apparent once the covariance is broken (see \cite{BC} for further details).
 
The expansion scalar $\Theta$ diverges positively at the initial time, indicating an initial singularity. As $t$ increases, $\Theta$ monotonically decreases, describing an expansion phase that progressively slows. Subsequently, during the phase where $\Theta$ becomes negative, the gravitational attraction dominates and a second gravitational singularity occurs. This is confirmed by the divergence of the expansion scalar $\Theta \to -\infty$ and the corresponding blow-up of the energy density $\mu$.
 
During this evolution, the system's anisotropy increases rapidly in the initial stages, then decreases until the expansion stops, and finally begins to grow again until the system reaches the final singularity. Furthermore, \ref{Fig:LRSIII1} shows that moving forward in time from the initial time, the weak energy condition remains preserved until the system reaches the future singularity. Nevertheless, moving toward the initial singularity, we find that the effective spinorial fluid violates the energy condition: the non-classical character of the fermion field becomes evident.

The specific solution analyzed here suggests that the role of the twist $\xi$ is significant only during the early stages of evolution and the geometry quickly approaches an LRSII space-time, indicating that the twist and fermion fields are not naturally compatible. This incompatibility is consistent with the analysis in \cite{VDFC}, where for $\eta=0$ LRSIII space-times were found to be inconsistent with the effective spinor fluid. In addition, the early-time dominance of $\xi$ also explains the violation of the weak energy condition: the terms proportional to $\xi$ in \eqref {thermo_LRSIII_np} make the pressures dominant, and this, in turn, induces $\mu+p_r$ to become negative.


\section{Conclusion}\label{Conclusion}
We employed the polar decomposition to express the Dirac field entirely in hydrodynamic terms, thereby avoiding the use of the tetrad formalism, the Dirac matrices and their specific representations. This enabled us to apply the powerful geometrical machinery of the covariant formalism to the study of a self-gravitating Dirac field in LRS space-times. More in particular, after performing the $(1+1+2)$ decomposition of both the spinor energy--momentum tensor and the Dirac equations in polar form, we investigated the Dirac field together with its gravitational backreaction in LRS geometries of types I, II, and III.

The present paper extends our previous work \cite{VDFC}, in which it appeared natural to identify the time-like and space-like congruences underlying the $(1+1+2)$ decomposition with the integral curves of the unit vector fields associated with velocity $u^i$ and spin $s^i$ of the Dirac field, respectively. However, this identification may be overly restrictive and could account for some of the obstructions encountered in our earlier analysis. A more general possibility, while remaining within the framework of LRS space-times, is to choose the tangent vectors to the congruences such that they lie pointwise in the plane spanned by the vector fields $u^i$ and $s^i$, without necessarily coinciding with them. This is precisely the approach adopted in the present work.

Within this more general setting, we found a broader class of solutions, mainly associated with the LRSIII case, which was previously precluded by requiring $u^i$ and $s^i$ to coincide with the generators of the temporal and spatial congruences \cite{VDFC}. By removing this identification, we showed that a self-gravitating Dirac field can fill an LRSIII space-time, provided that the associated spinorial fluid is not perfect. In the subsequent numerical analysis, we found that the role of the twist $\xi$ is significant only during the early stages of the evolution; the system then evolves toward an LRSII-like geometry, eventually reaching an anisotropic singularity. This behavior may suggest a possible tension between the space-time twist and the fermion field, indicating that these two elements might not be naturally compatible within a stable long-term dynamical evolution.

In the LRSII case, our results revealed a strong dependence of the space-time evolution on the initial data. Indeed, by modifying only the initial conditions for the expansion rate, we obtained two radically different dynamical behaviors. In both scenarios, the evolution originates from an initial singularity characterized by diverging expansion scalar $\Theta$ and shear $\Sigma$. We identified a first regime describing an anisotropic universe that initially expands and subsequently collapses into a final anisotropic singularity, and a second regime in which no collapse occurs, leading instead to a progressive isotropization of the space-time. In both cases, the dynamics is predominantly driven by the effective thermodynamic quantities, while the intrinsic behavior of the spinor field exerts only a limited influence on the evolution of the space-time geometry.

Among the explored scenarios, the case of stationary LRSI space-times emerged as particularly compelling. In these geometries, the effective fermion fluid appears to reproduce the physical conditions characteristic of the interior of compact matter distributions. On the one hand, this may suggest the intriguing possibility of a vortical relativistic star entirely sourced by a fermion field; on the other hand, it may provide a framework for analyzing fluid-like semiclassical models of nucleons.

However, while this conceptual framework is promising, several technical and theoretical aspects still require further investigation. A primary issue concerns the formulation of appropriate junction conditions; specifically, it remains unclear how the presence of the fermion field modifies the standard Israel matching conditions when joining an internal LRSI solution to an external vacuum space-time. Furthermore, the behavior of the surface gravity within the matter distribution presents a subtle challenge, as it appears to undergo a sign change that calls for a deeper physical interpretation. A rigorous analysis of these issues, requiring more specialized analytical tools and a dedicated treatment of boundary dynamics, will be the subject of forthcoming work.
\vspace{10pt}

\textbf{Data availability}. The manuscript does not have associated data in any repository.

\

\textbf{Conflict of interest}. There is no conflict of interest.

\end{document}